\documentclass[]{emulateapj}

\begin{document}

\title{A Spitzer survey of mid-infrared molecular emission from protoplanetary disks I: Detection rates}

\author{Klaus M. Pontoppidan\altaffilmark{1}}

\author{Colette Salyk\altaffilmark{2}}

\author{Geoffrey A. Blake\altaffilmark{1}}

\author{Rowin Meijerink\altaffilmark{1,3}}

\author{John S. Carr\altaffilmark{4}}

\author{Joan Najita\altaffilmark{5}}

\altaffiltext{1}{California Institute of Technology, Division of Geological and Planetary Sciences, 
MS 150-21, Pasadena, CA 91125, USA}
\altaffiltext{2}{The University of Texas at Austin, Department of Astronomy, 1 University Station C1400, Austin, Texas 78712, USA}
\altaffiltext{3}{Leiden Sterrewacht, P.O. Box 9513, NL-2300 RA, The Netherlands }
\altaffiltext{4}{Naval Research Laboratory, Code 7211, Washington, DC
  20375, USA}
\altaffiltext{5}{National Optical Astronomy Observatory, 950 N. Cherry
Ave. Tucson, AZ 85719, USA}

\begin{abstract}
We present a Spitzer InfraRed Spectrometer (IRS) search for 10-36\,$\mu$m molecular 
emission from a large sample of protoplanetary disks, including lines from H$_2$O, OH, C$_2$H$_2$, HCN and CO$_2$. 
This paper describes the sample and data processing and derives the detection rate of 
mid-infrared molecular emission as a function of stellar mass. The sample covers a 
range of spectral type from early M to A, and is supplemented by archival spectra of 
disks around A and B stars. It is drawn from a variety of nearby star forming regions, 
including Ophiuchus, Lupus and Chamaeleon. Spectra showing strong emission lines are 
used to identify which lines are the best tracers of various physical and chemical 
conditions within the disks.  In total, we identify 22 T Tauri stars with strong 
mid-infrared H$_2$O emission. Integrated water line luminosities, where water vapor is
detected, range from $5\times 10^{-4}$ to $9\times 10^{-3}\,\rm L_{\odot}$, likely making
water the dominant line coolant of inner disk surfaces in classical T Tauri stars.
None of the 5 transitional disks in the sample show detectable gaseous molecular emission 
with Spitzer upper limits at the 1\% level in terms of line-to-continuum ratios (apart from H$_2$), but the sample is too small to conclude whether this is a general property of transitional disks. 
We find a strong dependence on detection rate with spectral type; no disks around our sample of 25 A and B 
stars were found to exhibit water emission, down to 1-2\% line-to-continuum ratios, 
in the mid-infrared, while almost 2/3 of the disks around K stars show sufficiently intense water
emission to be detected by Spitzer. Some Herbig Ae/Be stars show tentative H$_2$O/OH emission 
features beyond 20\,$\mu$m at the 1-2\% level, however, and one of them shows CO$_2$ in emission. We argue that the observed differences 
between T Tauri disks and Herbig Ae/Be disks is due to a difference in excitation and/or 
chemistry depending on spectral type and suggest that photochemistry may be playing an
important role in the observable characteristics of mid-infrared
molecular line emission from protoplanetary disks.
  
\end{abstract}

\keywords{astrochemistry -- planetary systems: protoplanetary disks -- stars: pre-main sequence}

\section{Introduction}

Molecular volatiles, including H$_2$O, CO$_2$ and CH$_4$, among others, are thought to 
play a central role in the early evolution of planetary systems. Evidence collected
from Solar System material has shown that the Solar Nebula was 
characterized by a diverse chemistry during the process of planet formation. 
This chemistry played a pivotal role in the formation of giant planets by providing 
a reservoir of volatile solids to aid in a rapid build-up of planetary cores. 
It has long been thought that the volatiles that formed in the nebula 
seeded the surfaces and atmospheres of the terrestrial planets after their formation, 
the so-called ``late veneer''  model. The mechanism for delivering a veneer of water to an otherwise dry Earth is controversial, although primitive
carbonaceous chondrites provide the best match to the D/H ratio of the Earth's oceans
(for an overview, see \citealp{Morbidelli00}). 
Alternatively, it has been suggested that Earth's water was present at earlier stages, 
having been chemisorbed to small dust grains at temperatures too high for the formation 
of ices --- a scenario possible if the Earth-forming oligarchs formed in a disk 
environment rich in water vapor \citep{Muralidharan08}. 

Because these models are based on evidence gathered from ancient Solar System material,
very little is known about how volatiles evolve and influence planet formation 
in current analogs to the Solar Nebula --- the so-called protoplanetary disks. 
The material that formed the present-day Solar System originated in feeding zones 
located at distances within $\sim$20 AU from the Sun, known putatively as the 
{\it planet-forming region}.  It is a natural conjecture that solar nebula 
analogs can be found among the zoo of protoplanetary disks around young stars. 

Theory predicts that water plays a role in the physics of planet formation 
(in addition to the chemistry). Because O-, C- and N-based volatiles, 
with H$_2$O likely being the most abundant, account for half the condensible mass in 
a protoplanetary disk, the formation, transport and phase changes of these molecular 
species can profoundly affect the evolution of protoplanetary disks. 
First, theory predicts that the dynamic interplay between, in particular, water vapor 
and the buildup of planetesimals is crucial for planet formation. Processes such as 
freeze-out, gas diffusion and inwards gas-drag migration of solids may act to concentrate 
large amounts of ice and water vapor close to the midplane snow line, facilitating planet 
growth \citep[e.g.,][]{Stevenson88, Johansen07, Dodson-Robinson09, Ciesla09}. 
It is still an open question how the inward and outward water transport processes balance, 
but it seems likely that strong abundance gradients may be present in the inner 10 AU 
of protoplanetary disks, in both the radial and vertical directions. Such molecular abundance 
structures --- not only for water, but for most molecular species --- should be observable. 

The planet-forming region represents a chemical environment very different than 
those studied in the interstellar medium, molecular clouds and even the outer regions 
($>$100\,AU) of the same disks. The very high densities, temperatures, radiation fields and,
importantly, short dynamical and chemical time scales in such regions generate conditions 
that may be more easily compared to the chemistry of planetary atmospheres \citep[e.g.,][]{Woitke09}. 
In direct comparison, models for planetary atmospheres show that complexities are 
high and predictabilities low, making the acquisition of empirical data crucial 
for understanding inner disk chemistry and disk ``climate''. 

The recent detections of a large number of rotational emission lines due to warm water vapor, 
in addition to the rovibrational Q-branches of HCN, C$_2$H$_2$ and CO$_2$ in three protoplanetary 
disks --- AA Tau, AS205N and DR Tau \citep{Carr08,Salyk08} --- have prompted a number of questions: 
How common is this emission? If a large number of protoplanetary disks exhibit strong molecular 
emission in the mid-infrared, the observed lines have the potential to be a unique tracer of 
inner disk chemistry as well as disk evolution and planet formation processes. 
Are differences from disk to disk present? If many disks exhibit similar molecular signatures, 
they will serve as good tracers of physical and chemical conditions or disk evolution. A fundamental 
question is therefore which physical and chemical processes result in the presence or absence of 
molecular emission tracers. If lines are absent in some disks, for example, is it due to abundance or 
excitation differences? 

We have collected a database of high quality Spitzer spectra aimed at the analysis of gas-phase
emission lines, based on both new, dedicated observations, as well as on re-processed archival data. Given
the large amount of information contained in each spectrum, we anticipate presenting a series of
papers focusing on different aspects of the data. 
In this first manuscript, we present the data and use a sample of $\sim$75 protoplanetary disks spanning
a wide range in stellar effective temperature to demonstrate that the presence of complex molecular emission, 
including that of water vapor, is indeed a common property 
of protoplanetary disks around stars of solar mass and lighter. We also show that
disks around A and B stars do not show strong molecular emission, as defined by having line-to-continuum
ratios lower than their T Tauri counterparts by an order of magnitude or more. A companion paper presents a detailed 
analysis of the molecular emission spectra, including a comparison to emission models and a detailed study
of the full range of molecular features, including HCN, C$_2$H$_2$, CO$_2$ and OH features (Salyk et al. 2010, herafter Paper II). 
Future work will describe a companion study of T  Tauri disks of the Taurus star forming region (Carr et al., Najita et al., in prep), as well as
more detailed modeling and comparative studies of stellar properties such as UV and X-ray fluxes, accretion rates, etc. 

\section{Observations}

\subsection{Sample selection}
The young star+disk systems examined in this paper are drawn from several Spitzer-IRS 
short-high (SH) and long-high (LH) observing programs. Most of the Chamaeleon and Lupus sources
were observed as part of a deep T Tauri star survey (PID 50641, PI. J. Carr, hereinafter GO-5) that was
designed to obtain very high S/N spectra by maximizing 
redundancy and using dedicated background observations for each target. The subset of the GO-5 sample presented here 
was constructed by selecting disks spanning the known range in disk infrared colors and stellar
X-ray luminosity, in particular those previously observed in the the cores to disks (c2d) Legacy 
program \citep{Evans03} that showed some hint of water emission (but which were not observed 
with sufficient redundancy and high enough S/N to produce firm detections). 
Another subset of targets from the high quality survey focused on the Taurus cluster, and 
will be the subject of a separate study (Carr et al., in prep). The sample covered 
here is summarized in Tables \ref{TT_table} and \ref{HAeBe_table}.

The non-Taurus `high-quality' sample is supplemented by additional archival c2d spectra; 
and for comparison with the T Tauri stars that dominate the c2d selection, spectra from the Herbig Ae/Be star 
survey by J. Bouwman \citep[PID 3470][]{Boersma08} were extracted from the Spitzer archive 
and reduced using the same procedure as the remaining sample. This ensures that the sample
includes a significant number of disks associated with stars of spectral types spanning 
from $M$ to $B$. The full sample includes disks from the young clusters in Perseus, Taurus, 
Chamaeleon, Lupus, Ophiuchus and Serpens, but is not complete in any strict sense. For instance, 
the sample represents a selection of the ``best'' available Spitzer spectra, and thus tends 
to be biased toward brighter, isolated, disks with low background emission \citep[the c2d selection criteria are described in ][]{Evans03}. Nevertheless, 
given the size of the sample (75 disks),  relative to the total number of class II disks 
brighter than $\sim$100 mJy at 8\,$\mu$m in the major nearby star forming clouds surveyed by
the c2d program \citep[$\sim 120$,][]{Evans09}, there is little room for strong biases, except 
for those imposed by the IRS sensitivity limit. A disk with a flat spectrum and a flux of 
100 mJy at 8\,$\mu$m roughly corresponds to a star with mass $\sim$0.1 M$_{\odot}$ at the 
distance of Ophiuchus (125 pc), and 0.5\,$M_{\odot}$ at the distance of Serpens \citep[415 pc][]{Dzib10}, using 
the evolutionary tracks of \cite{Siess00}. 

\begin{figure}
\centering
\includegraphics[width=8cm]{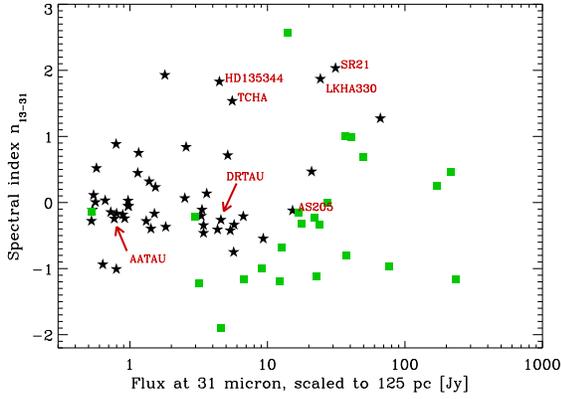}
\caption[]{Distribution of mid-infrared spectral indices and 31 $\mu$m fluxes of the sample. 
The stars indicate T Tauri stars (spectral type $\le$F), while
the squares denote Herbig Ae/Be stars.
}
\label{spindex}
\end{figure}

Figure \ref{spindex} shows the distribution of spectral indices $n_{13\,\mu m - 31\,\mu m}$ 
of the sample, as defined in \cite{Kessler06} and \cite{Furlan06}.
A few sources are labeled, for reference, including the transitional disks T Cha, LkHa 330, 
HD 135344B and SR 21. 
It can be seen that the sample spans the range from disks with 
slopes that are nearly photospheric, $n_{\rm 13\,\mu m - 31\,\mu m}$$\sim$-2,
to transitional disks with large inner holes and rising fluxes into the far-IR
$n_{\rm 13\,\mu m - 31\,\mu m}$$\sim$2. The majority of the
disks center around $n_{\rm 13\,\mu m - 31\,\mu m}$$\sim$-1 -- 0, a distribution
quite similar to that of the general disk populations of the included star forming 
regions \citep{Furlan09}. 
The Herbig Ae/Be stars tend to be brighter than the T Tauri stars, but not enough to reflect 
their much higher luminosities -- an
indication that the Herbig Ae/Be stars are, on average, located at larger distances than 
the T Tauri stars. The sample also shows a deficit of transitional disks with low 
31\,$\mu$m flux.   This may, to some extent, reflect 
an intrinsic property of transitional disks,  although one canonical transitional
disk that is not in our sample --- GM Aur --- only has a $31\,\mu$m brightness of $\sim1$ Jy \citep{Furlan06}.

\begin{deluxetable*}{llllllll}

\tablecaption{Sample of protoplanetary disks around T Tauri stars ($\leq$F)}

\tablehead{
\colhead{Source name}   & \colhead{Sp. Type} & \colhead{Distance} & \colhead{SH int. time} & \colhead{LH int. time} & \colhead{BG obs.\tablenotemark{a}} & \colhead{AOR}  & \colhead{Obs. program} \\
                                       &                               &\colhead{[pc]}           & \colhead{seconds}      & \colhead{seconds}       &                                                          &                         &
}
\startdata
LkHa 270        &K7$^3$           & 250 &$4\times31.5  $& $4\times61   $ &2&r5634048& c2d\\ 
LkHa 271        &K3$^3$           & 250 &$4\times122   $& $4\times61   $ &2&r11827968& c2d\\ 
LkHa 326        &M0$^4$          & 250 &$48\times31.5 $& $28\times61  $ &1&r27063552& 50641\\
LkHa 327        &K2$^{2,3}$       & 250 &$4\times6.3   $& $4\times14.7 $ &2&r5634560& c2d\\
LkHa 330        &G3$^{2,3}$       & 250 &$28\times31.5 $& $32\times14.7$ &1&r27063040& 50641\\
LkCa 8            &M0$^5$          & 140 &$16\times31.5 $& $4\times61   $ &2&r9832960& c2d\\  
IQ Tau            &M0.5$^5$        & 140 &$4\times31.5  $& $4\times61   $ &2&r9832704& c2d\\  
V710 Tau       &M1/M3$^5$     & 140 &$4\times122   $& $8\times61   $ &2&r5636608& c2d\\ 
AA Tau           &K7$^5$            & 140 &$48\times31.5 $& $32\times61  $ &1&r14551552& 20363\tablenotemark{b}\\ 
CoKu Tau/4   &M1.5$^5$         & 140 &$4\times31.5  $& $8\times61   $ &2&r5637888& c2d\\ 
DN Tau          &M0$^5$            & 140 &$8\times31.5  $& $4\times61   $ &2&r9831936& c2d\\ 
FX Tau           &M1/M2$^{5,6}$  & 140 &$4\times31.5  $& $4\times61   $ &2&r9832448& c2d\\ 
DR Tau          &K7$^5$              &140 &$48\times6.3  $& $40\times14.7$ &1&r27067136& 50641\\ 
SX Cha           &M0$^{1,7}$         &180 &$28\times31.5 $& $32\times61  $ &1&r27066624& 50641\\
SY Cha           &M0$^{1,7}$         &180 &$56\times31.5 $& $48\times61  $ &1&r27066368& 50641\\ 
TW Cha          &K7$^1$              &180 &$56\times31.5 $& $48\times61  $ &1&r27066368& 50641 \\  
VW Cha          &M0.5$^1$          &180 &$24\times31.5 $& $24\times61  $ &1&r27066112& 50641\\  
VZ Cha           &K6$^1$             & 180 &$32\times31.5 $& $48\times61  $ &1&r27065856& 50641\\  
WX Cha           &M0$^7$            & 180 &$48\times31.5 $& $48\times61  $ &1&r27065600& 50641 \\ 
XX Cha            &M1$^7$            & 180 &$56\times31.5 $& $48\times61  $ &1&r27065344& 50641\\  
T Cha              &G8$^{8,9}$         & 66 &$4\times31.5  $& $2\times61   $ &2&r5641216& c2d\\ 
Sz 50              &M3$^{10}$         & 180 &$56\times31.5 $& $48\times61  $ &1&r27065088& 50641\\  
HD 135344B   &F8$^{11}$          & 84 &$2\times122   $& $4\times61   $ &2&r5657088& c2d\\ 
HT Lup           &K2$^{13}$          & 150 &$40\times6.3  $& $32\times14.7$ &1&r27064832& 50641\\   
GW Lup           &M0$^1$            & 150 &$4\times122   $& $8\times61   $ &2&r5643520& c2d\\ 
GQ Lup           &K7$^{14}$          & 150 &$32\times31.5 $& $24\times61  $ &1&r27064576& 50641\\ 
IM Lup            &M0$^{15}$         & 150 &$24\times31.5 $& $24\times61  $ &1&r27064320& 50641\\ 
HD 142527    &F6-F9$^{8,9}$    & 150 &$4\times6.3$  &$4\times6.3 $  &1&r11005696 &  3470\\ 
RU Lup           &K7$^{2,12,13}$     & 150 &$20\times31.5 $& $20\times31.5$ &1&r27064064& 50641\\ 
RY Lup           &K0-G0$^{1,13}$   & 150 &$2\times31.5  $& $2\times61   $ &2&r5644544& c2d\\ 
EX Lup           &M0$^1$              & 150 &$24\times31.5 $& $24\times61  $ &1&r27063808& 50641\\ 
AS 205          &M3-K0/K5$^{16,20}$ &  125   &$4\times6.3   $& $24\times14.7$ &2/1&r5646080/r27063296& c2d/50641\\ 
Haro 1-1       &K5$^{17}$           & 125 &$8\times31.5  $& $2\times61   $ &2&r9833472& c2d\\  
Haro 1-4       &K6-7$^1$          & 125 &$4\times31.5  $& $2\times61   $ &2&r9833216& c2d\\
VSSG1           &K7$^{18}$           & 125 &$2\times31.   $& $4\times14.7 $ &2&r5647616& c2d\\
DoAr 24E      &G6-K0$^{19,20}$  & 125&$20\times31.5 $& $32\times14.7$ &1&r27062784& 50641\\
DoAr 25        &K5$^{19}$            & 125&$4\times122   $& $4\times61   $ &2&r12663808& c2d\\ 
SR 21            &G2.5$^{20}$         & 125&$2\times31.5  $& $4\times14.7 $ &2&r5647616& c2d\\     
SR 9              &K5$^{19}$            & 125&$2\times31.5  $& $4\times61   $ &2&r12027392& c2d\\      
V853 Oph     &M4$^{19}$            &125&$8\times31.5  $& $8\times61   $ &2&r12408576& c2d\\ 
ROX 42C       &K4/K5$^{21}$       &125&$4\times31.5  $& $2\times61   $ &2&r6369792& c2d\\ 
ROX 43A       &G0$^{22}$            & 125&$2\times31.5  $& $4\times61   $ &2&r15914496& c2d\\ 
Haro 1-16    &K2-3$^1$           & 125&$28\times31.5 $& $24\times61  $ &1&r27062016& 50641\\
Haro 1-17    &M2.5$^{23}$        & 125&$8\times122   $& $8\times61   $ &2&r11827712& c2d\\ 
RNO 90         &G5$^2$                & 125&$36\times6.3  $& $24\times14.7$ &1&r27061760& 50641\\   
Wa Oph/6     &K$^{24}$              & 125&$24\times31.5 $& $24\times61  $ &1&r27060736& 50641\\
V1121Oph   &K4$^{25}$             & 125&$2\times31.5  $& $2\times61   $ &2&r5650688& c2d\\ 
EC 82           &M0$^{17}$           &  415&$32\times31.5 $& $24\times61  $ &1&r27059712& 50641\\     
\enddata

\tablecomments{REFERENCES -- (1) \cite{Guenther07}; (2) \cite{Levreault88}; (3) \cite{Fernandez95}; (4) \cite{Casali96}; (5) \cite{Kenyon95}; (6) \cite{Nguyen09}; 
(7) \cite{Hartmann98} (8) \cite{Alcala93}; (9) \cite{Schisano09}; (10) \cite{Hughes92}; (11) \cite{Coulson95}; (12) \cite{Stempels07}; (13) \cite{Stempels03}; 
(14) \cite{Duarte08}; (15) \cite{Hughes94}; (16) \cite{Cohen79}; (17) \cite{Martin98}; (18) \cite{Natta06}; (19) \cite{Wilking05}; (20) \cite{Prato03};
(21) \cite{Lee94}; (22) \cite{Bouvier92}; (23) \cite{Rydgren82}; (24) \cite{Grankin07}; (25) \cite{Torres04}; (26) \cite{Stephenson77}}

\tablenotetext{a}{Type of background observation: 1 -- dedicated background observation obtained at the same time at the on-source observation. 2 -- ``best effort'' archival background observation.}
\tablenotetext{b}{From \cite{Carr08}}

\label{TT_table}
\end{deluxetable*}

\begin{deluxetable*}{llllllll}
\tablecaption{Sample of protoplanetary disks around Herbig Ae/Be stars ($\geq$A)}
\tablehead{
\colhead{Source name}   & \colhead{Sp. Type} & \colhead{Distance}& \colhead{SH int. time}& \colhead{LH int. time} & \colhead{BG obs.\tablenotemark{a}} & \colhead{AOR}  & \colhead{Obs. program}\\
                                       &                               &\colhead{[pc]}           & \colhead{seconds}      & \colhead{seconds}       &                                                          &                         &
}
\startdata
HD 36112  &A3-A5$^{1,8}$    & 200&$4\times6.3$  &$4\times6.3 $  &1&r11001088 &  3470\\ 
HD 244604 &A0$^{1,8}$          &400&$6\times6.3$  &$8\times14.7$  &1&r11001344 &  3470\\ 
HD 36917  &B9.5/A0.5$^{2,8}$&400&$4\times6.3$  &$6\times14.7$  &1&r11001600 &  3470\\ 
HD 37258  &A1-A2$^{3,8}$    & 506&$8\times6.3$  &$4\times61  $  &1&r10998784 &  3470 \\ 
BF Ori    &A5$^4$                   & 400&$2\times31.5$ &$2\times61  $  &2&r5638144  &  c2d\\
HD 37357  &A0$^{5,8}$           & 506 &$8\times6.3$  &$6\times14.7$  &1&r11001856 &  3470\\ 
HD 37411  &B9$^{5,8}$           & 506 &$6\times31.5$ &$8\times14.7$  &1&r11002112 &  3470\\ 
RR Tau    &A0$^{6}$                & 2000&$2\times31.5$ &$2\times61  $  &2&r5638400  &  c2d\\
HD 37806  &B9-A2$^{1,8}$     & 473&$4\times6.3$  &$6\times6.3 $  &1&r11002368 &  3470\\ 
HD 38087  &B5$^{8}$             & 473&$4\times31.5$ &$8\times14.7$  &1&r11002624 &  3470\\ 
HD 38120  &B9$^{8}$             & 506&$4\times6.3$  &$4\times6.3 $  &1&r11002880 &  3470\\  
HD 50138  &B8$^{8}$             & 290&$4\times6.3$  &$4\times6.3 $  &1&r11003648 &  3470\\ 
HD 72106  &A0$^{8}$             &290&$8\times6.3$  &$6\times14.7$  &1&r11004416 &  3470\\ 
HD 95881  &A1$^{8}$             &118&$4\times6.3$  &$8\times6.3 $  &1&r11004928 &  3470\\ 
HD 98922  &B9$^{8}$              &1000&$2\times6.3$  &$2\times14.7$  &2&r5640704  &  c2d\\
HD 101412 &A0$^{8}$             &160&$2\times31.5$ &$2\times61  $  &2&r5640960  &  c2d\\
HD 144668 &A7$^{8}$             &208&$4\times6.3$  &$4\times6.3 $  &1&r11005952 &  3470\\ 
HD 149914 &B9.5$^{8}$           &165&$6\times31.5$ &$16\times61 $  &1&r11000832 &  3470\\ 
HD 150193 &A0$^{8}$              &150&$4\times6.3$  &$4\times6.3 $  &1&r11006208 &  3470\\ 
VV Ser    &A0-B6$^{6,7}$           & 415&$2\times31.5$ &$2\times61  $  &2&r5651200  &  c2d\\ 
LkHa 348     &B1$^{10}$            & 415&$4\times6.3   $& $4\times14.7 $ &2&r9831424& c2d\\
HD 163296 &A0$^{8}$              & 122&$4\times6.3$  &$4\times14.7$  &2&r5650944  &  c2d\\%
HD 179218 &B9$^{8}$               & 244&$4\times6.3$  &$4\times6.3 $  &1&r11006976 &  3470\\
HD 190073 &A0$^{8}$              & 767&$4\times6.3$  &$8\times6.3 $  &1&r11007232 &  3470\\
LkHa 224  &A4-F9$^{1,7}$          &980&$8\times6.3$  &$8\times14.7$  &2&r16827648 &  c2d\\ 
\enddata
\tablecomments{REFERENCES -- (1) \cite{Manoj06}; (2) \cite{Manoj02}; (3) \cite{Gray93}; (4) \cite{Grady96}; (5) \cite{The94}; (6) \cite{Hernandez04}; (7) \cite{Mora01}; (8) \cite{Oudmaijer92};
(9) \cite{Nordstrom04}; (10) \cite{Stephenson77}}

\tablenotetext{a}{Type of background observation: 1 -- dedicated background observation obtained at the same time at the on-source observation. 2 -- ``best effort'' archival background observation.}

\label{HAeBe_table}
\end{deluxetable*}

\subsection{Data reduction}
\begin{figure}
\centering
\includegraphics[width=8cm]{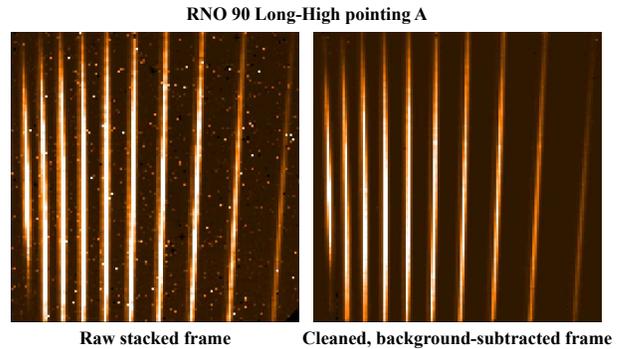}
\caption[]{Quality of the Long-High (LH) cleaning procedure and background subtraction. 
Dedicated background frames as well as high frame redundancy are especially critical for
high SNR LH observations, by enabling an efficient removal of bad pixels. }
\label{LH_clean}
\end{figure}

The Spitzer-IRS spectra are reduced using IDL scripts optimized for deep integrations 
with high redundancy, i.e., up to 56 individual spectral frames for our data set, and dedicated background 
observations. The procedure is similar to that used by \cite{Carr08} for AA Tau, and the 
resulting spectra have been benchmarked to those results. The reduction begins with the 
droop frames. On-source spectra as well as background exposures are co-added.  To avoid 
inappropriately weighting pixels at the edges of the entrance slit, the flat field is 
divided by low-order polynomial fits in both the dispersion and cross-dispersion directions. 
This produces a flat field that corrects the pixel-to-pixel response only. The background 
exposure is used to detect rogue pixels by flagging pixels with values that are more than 2$\sigma$ from 
the mean of that pixel. Next, the background frame is subtracted from the on-source frames 
(two nod positions) and the bad pixels are linearly interpolated in the dispersion direction. 
The ability to detect transient bad pixels by using a highly redundant (i.e. including many 
individual readouts) background observation taken essentially at the same time as the pointed 
observation produces excellent results. An example LH frame, before and after bad pixel cleaning, 
is shown in Figure \ref{LH_clean}. 

Spectra are extracted from the 2D co-added spectral images using optimal extraction \citep{Horne86}. 
A selection of 15 standard star observations of $\delta$ Dra and $\xi$ Dra were retrieved 
from the archive and reduced in the same manner, with the same flat field and extraction 
apertures, essentially producing a data base of spectral response functions (SRFs). Because 
the spectral response function depends on the location of the target in the slit, a data base 
of SRFs allows a selection of those that best match a given science observation. The best 6-12 standard stars
were chosen by minimizing the noise at 10-11 $\mu$m (SH orders 19 and 20) and a flux 
calibrated spectrum is produced for each standard star observation. The flux calibrated spectra 
are carefully defringed using IRSFRINGE \citep{Lahuis07} ensuring that the real structure from 
the complex water emission spectrum is not affected. Setting the tolerance of a defringer too 
low will essentially result in the data being processed by a low-pass filter, removing high 
frequency structure, such as that produced by the densely packed water spectrum. The behavior of IRSFRINGE
was tested by visually comparing the final product with spectra that were not defringed. In practice, 
most orders required no defringing at all, with most of the fringing pattern being removed by
the SRFs. The orders are then combined using a weighted average with the blaze function 
as weights. No additional relative scaling of the orders is necessary, as the match is generally 
within the noise. Finally, the set of standard star divided spectra that contain the least amount 
of residual fringes are co-added to produce a final spectrum. 

For the set of spectra that do not have dedicated background observations, such as the suite of 
c2d spectra, a background observation taken as close in time as possible ($<$6 months, typically 1-3 months) is used instead. Because 
no dedicated background observations were obtained for the c2d spectra, a few sources suffer
from an incomplete background subtraction. However, given the spatial undersampling of Spitzer 
high resolution spectra, it is difficult to subtract the background without introducing strong artifacts. 

\begin{figure}
\centering
\includegraphics[width=8cm]{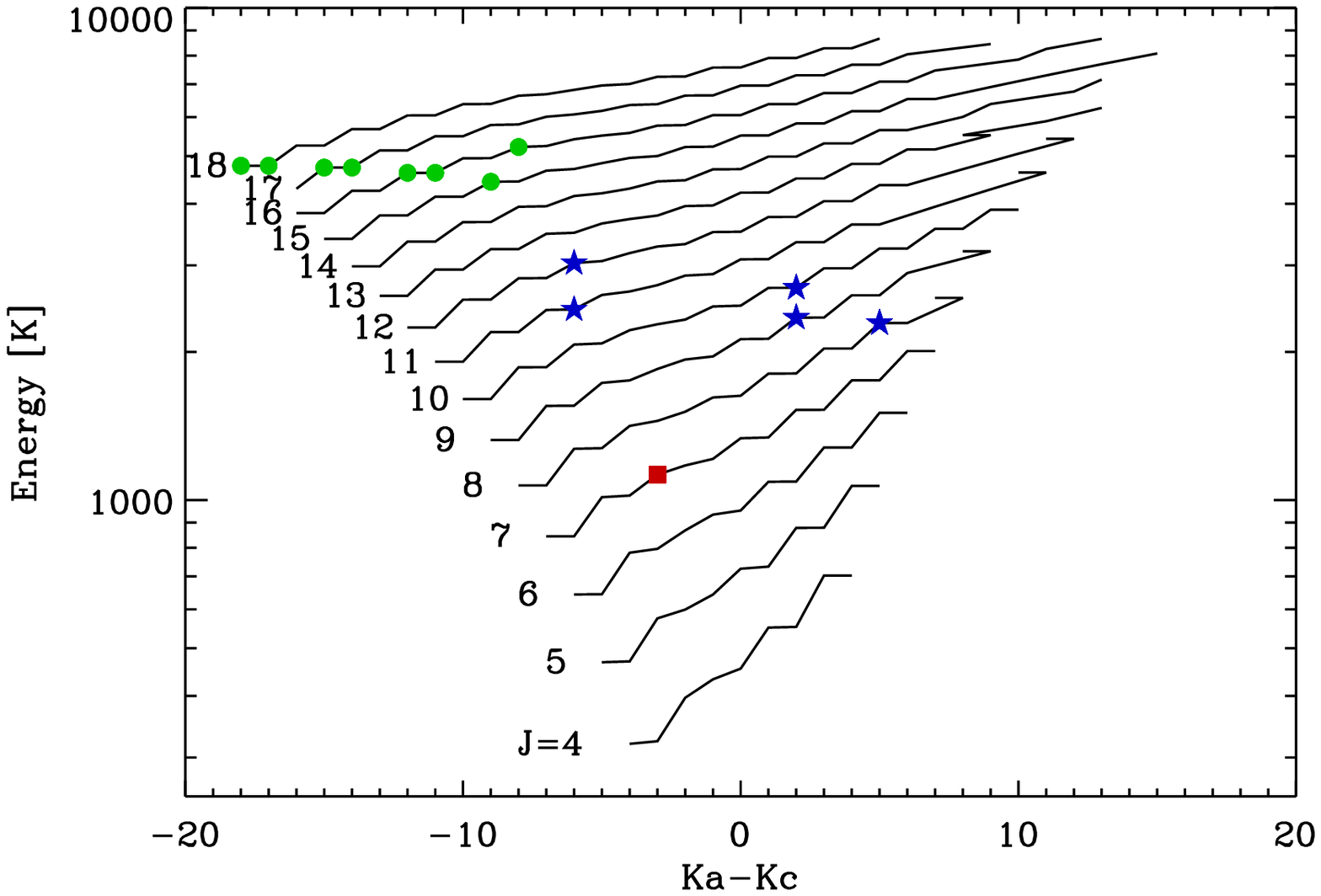}
\caption[]{Energy diagram of the rotational ladder of H$_2^{16}$O. The numbers indicate
the $J$ rotational quantum number. Note how the energies of the levels generally increase
within a $J$ ladder with $K_a-K_c$. Green circles indicate the dominant
upper levels of the 13.3 and 29.5\,$\mu$m line complexes. Blue stars indicate the dominant upper levels of the 15.17 and 17.22\,$\mu$m
complexes. The red square is the dominant upper level of the 29.85\,$\mu$m line. }
\label{H2O_rotladder}
\end{figure}

\section{The mid-infrared molecular spectrum of a typical protoplanetary disk}

\begin{figure*}
\centering
\includegraphics[width=16cm]{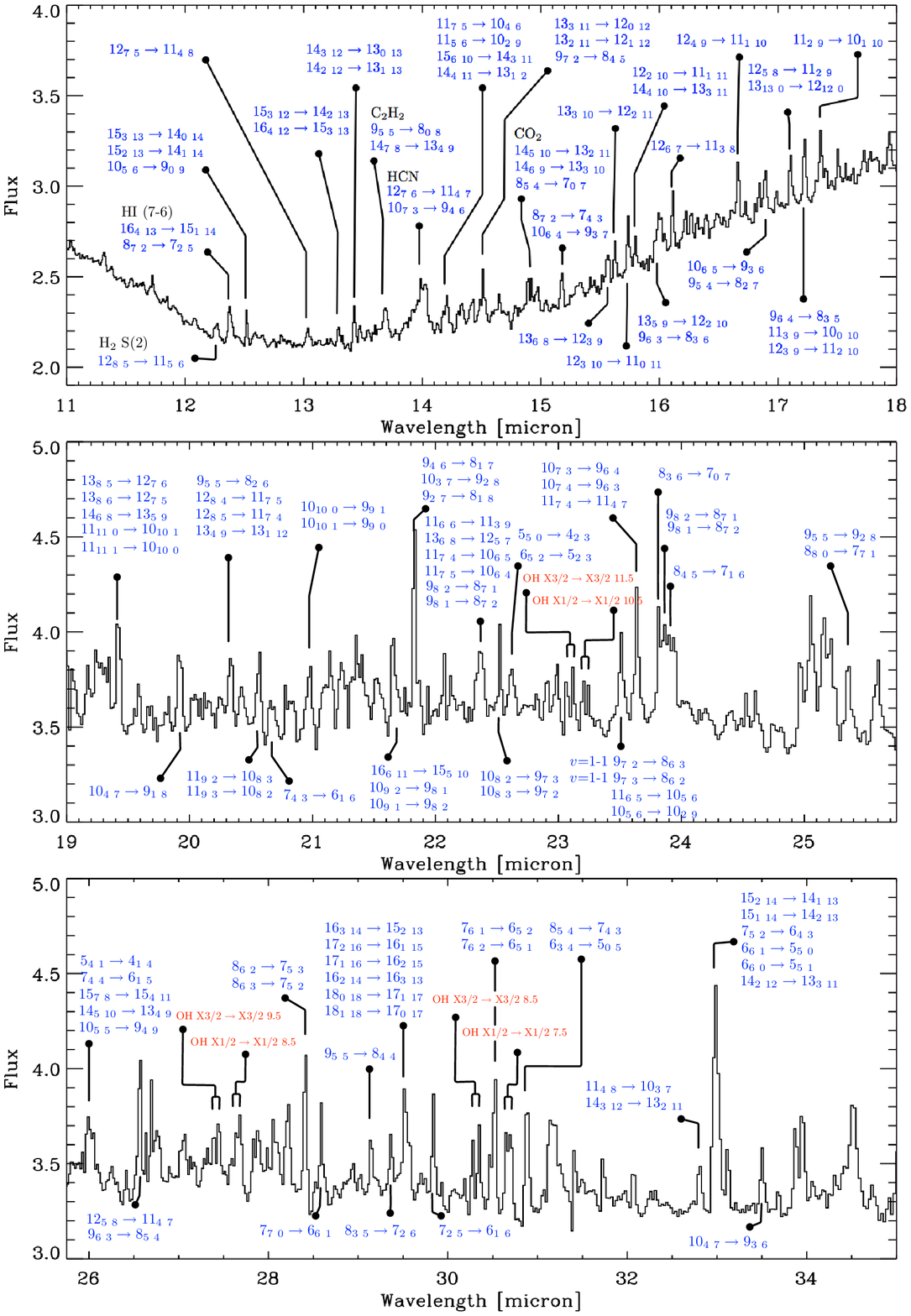}
\caption[]{Representative identification, using the spectrum of RNO 90, of selected major molecular line complexes in the Spitzer high resolution spectral range (many
features are left unlabeled for clarity). Unless otherwise noted, the transitions
refer to the rotational quantum numbers $J_{Ka\ Kc}$ in the ground vibrational state of H$_2^{16}$O.  Rotational transitions in
the vibrational ground state where $Ka+Kc$ is odd have ortho nuclear spin functions.}
\label{line_ident}
\end{figure*}

The molecular emission spectrum from an optically thick protoplanetary disk, where detected, is characterized 
by $>$100 line complexes present throughout the Spitzer high resolution 10 - 36 $\mu$m wavelength range.
The vast majority of lines are due to rotational transitions of the main 
water isotopologue, H$_2^{16}$O. Recall that this asymmetric top molecule gives rise to three 
rotational quantum numbers. We use the HITRAN catalog \citep{Rothman05} for line identification 
and adopt the standard notation, such that a 
rotational level is defined by $J_{Ka\, Kc}$. In general, the level energy increases, within 
a $J$ state, with the difference $K_a-K_c$, with a few minor exceptions. This is illustrated 
in Figure \ref{H2O_rotladder}, which shows the energy levels as a function of $K_a-K_c$. 

\begin{figure}
\centering
\includegraphics[width=8cm]{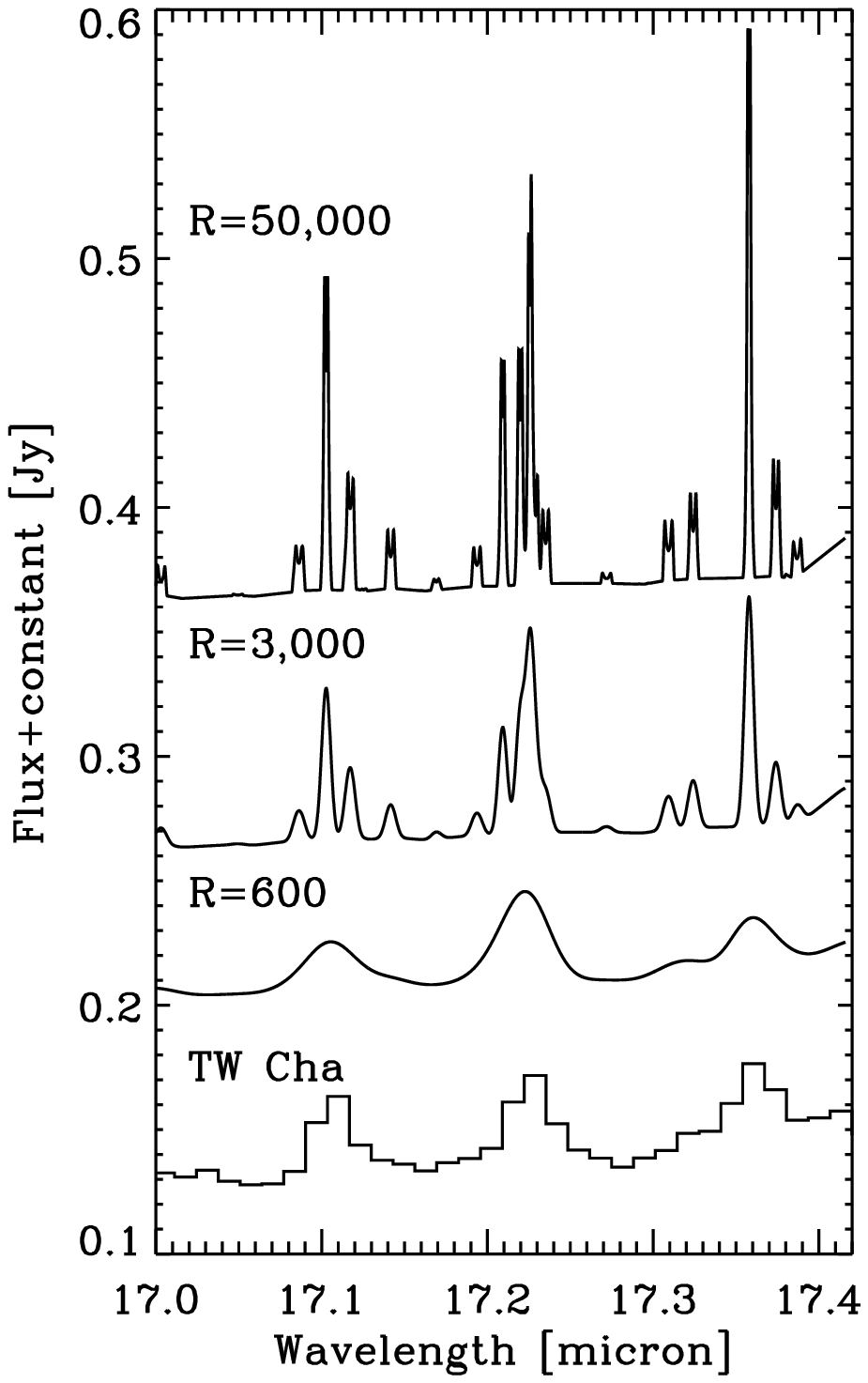}
\caption[]{ LTE disk model showing how the 17.22\,$\mu$m H$_2$O line complex breaks up into many individual lines at higher spectral resolution. 
Model spectra with resolving powers relevant for both Spitzer-IRS ($R=600$) and JWST-MIRI (R=3000) are shown. Note how JWST-MIRI is
expected to resolve the water line complexes (but not the individual lines). The lower curve is the spectrum of TW Cha, shown for 
comparison. The model is taken from \cite{Pontoppidan09}. }
\label{highres_complex}
\end{figure}

Figure \ref{line_ident} shows the Spitzer SH/LH spectrum of RNO 90, a strong water emission source, with 
a selection of the strongest line complexes identified. The energy traced by each transition can be easily determined by 
inspection of Figure \ref{H2O_rotladder}. As can be seen in Figure \ref{line_ident}, most ``lines'' 
are in fact blends of 2-5 pure rotational water transitions, usually with a relatively wide range 
of excitation energies \citep[see also][]{Meijerink09}. The typical structure of a water line complex is illustrated in Figure \ref{highres_complex}. The 
identifications include transitions that dominate the total line blend flux, weaker lines are 
excluded for clarity. The Spitzer range includes transitions from $J=5$ to 
$J=18$; higher $J$'s are present, but at a weaker level. Many ``ortho/para'' transition pairs, 
namely those that come from upper levels with $K_a$ close to $J$, tend to lie very close
in frequency, causing them to be blended at R=600. One example 
of this is the ortho $10_{10\,1}\rightarrow 9_{9\,0}$ and para $10_{10\,0}\rightarrow 9_{9\,1}$ 
transitions at 20.97\,$\mu$m. This illustrates the importance of future high spectral resolution 
observations in determining the ortho/para ratio(s) in disks.  

As much as 30\% of the wavelength space is blanketed in water lines above the 1\% line-to-continuum 
level \citep{Pontoppidan09}. An important consequence of this is that peaks tend to appear in the Spitzer spectrum at places 
where lines crowd together, and not necessarily where a single, particularly strong line is present. 
RNO 90 is chosen as an illustrative example because it has one of the highest line-to-continuum 
ratios in the Spitzer sample, but it may not have particularly high contrast lines at higher 
resolving power. 

In addition to water, lines due to other molecular species are present in many Spitzer spectra, 
including that of RNO 90. OH exhibits a characteristic doublet pattern, and at least three 
sets of OH lines are visible around 23.2, 27.5 and 30.5\,$\mu$m. The Q branches of C$_2$H$_2$, 
HCN and CO$_2$ are seen at 13.7, 14.0 and 14.95\,$\mu$m, respectively, as also noted in \cite{Salyk08}, 
\cite{Carr08} and \cite{Pascucci09}. Additional features of note are the [Ne II] line at 12.814\,$\mu$m, 
the H$_2$ S(2) line at 12.279\,$\mu$m, as well as several H I lines across the Spitzer range, 
the brightest one being the (7-6) line 
at 12.371\,$\mu$m. In general, all of these transitions are blended with water lines, which 
should be taken into account when measuring line strengths in water-rich spectra. 

The water and organics line emission is spatially unresolved, in contrast to the H$_2$ lines 
which are sometimes extended \citep{Lahuis07}. This is confirmed both by inspection of the 
spectral images, as well as from the dedicated background observations, where available. 

\begin{figure*}
\centering
\includegraphics[width=7cm]{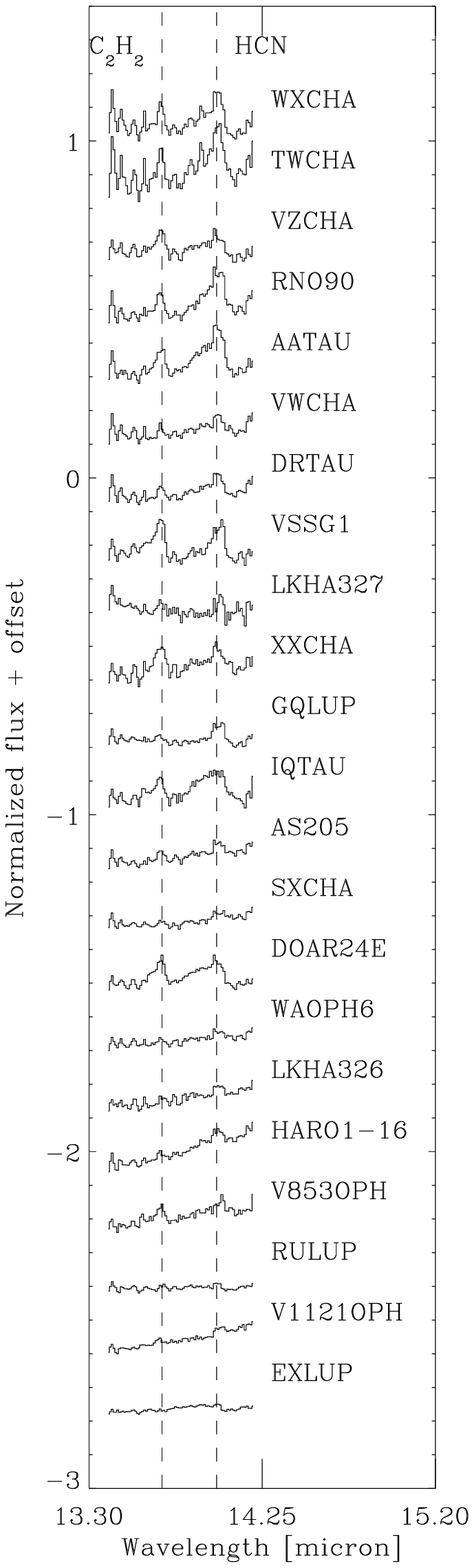}
\includegraphics[width=3.5cm]{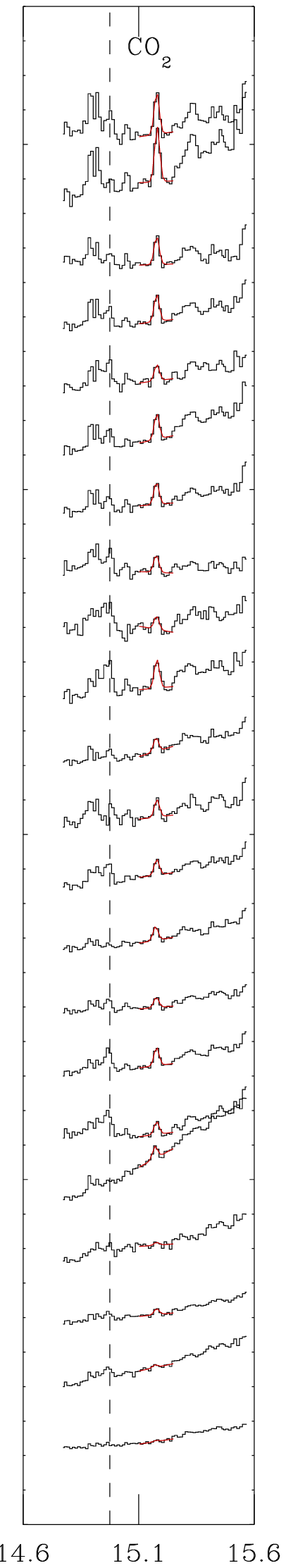}
\includegraphics[width=3.5cm]{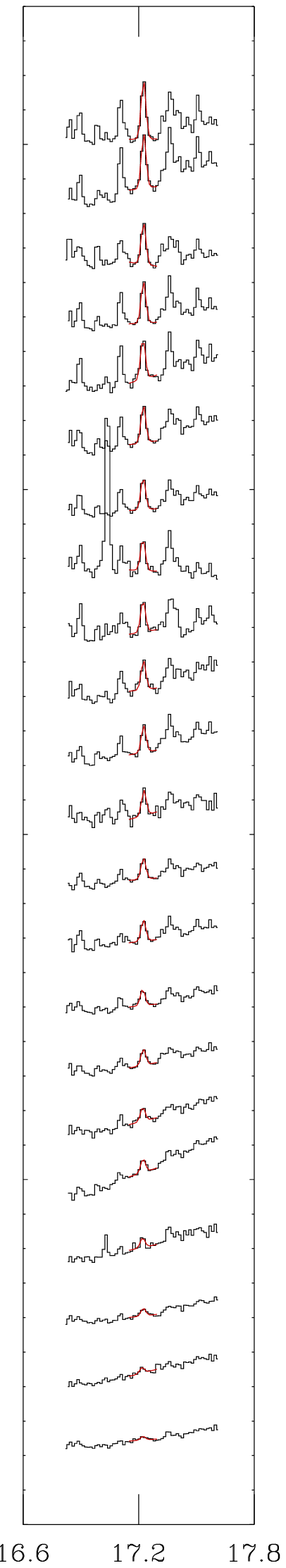}
\caption[]{Selected regions of the Spitzer-IRS SH module of T Tauri
  stars with detected H$_2$O. From left to right, the spectra are 
centered on: 1) the acetylene (C$_2$H$_2$) and HCN $Q$-branches, 2) the 15.17\,$\mu$m H$_2$O 
line complex, and 3) the 17.22\,$\mu$m H$_2$O line complex. The lines are marked with a Gaussian 
fit (red curves). The spectra are in order of decreasing line-to-continuum ratio of the 
17.22\,$\mu$m complex.}
\label{detectSH}
\end{figure*}

\begin{figure*}
\centering
\includegraphics[width=7cm]{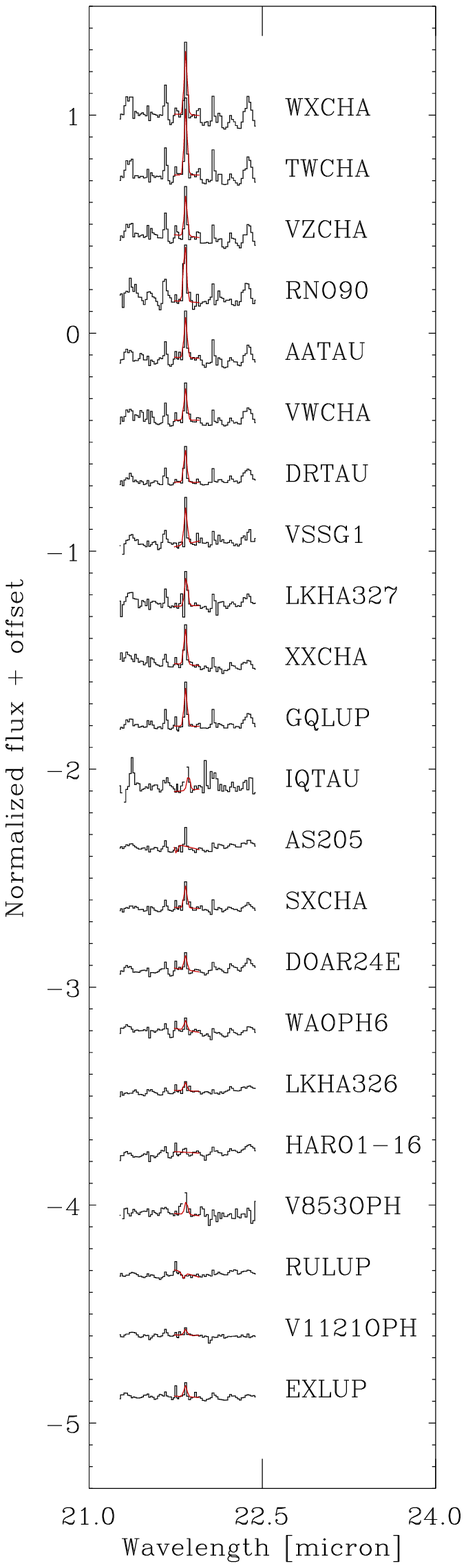}
\includegraphics[width=3.5cm]{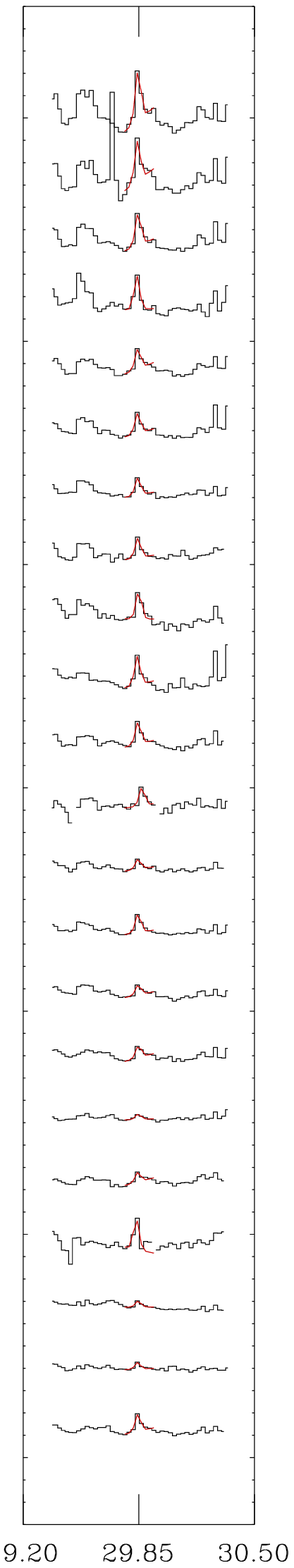}
\includegraphics[width=3.5cm]{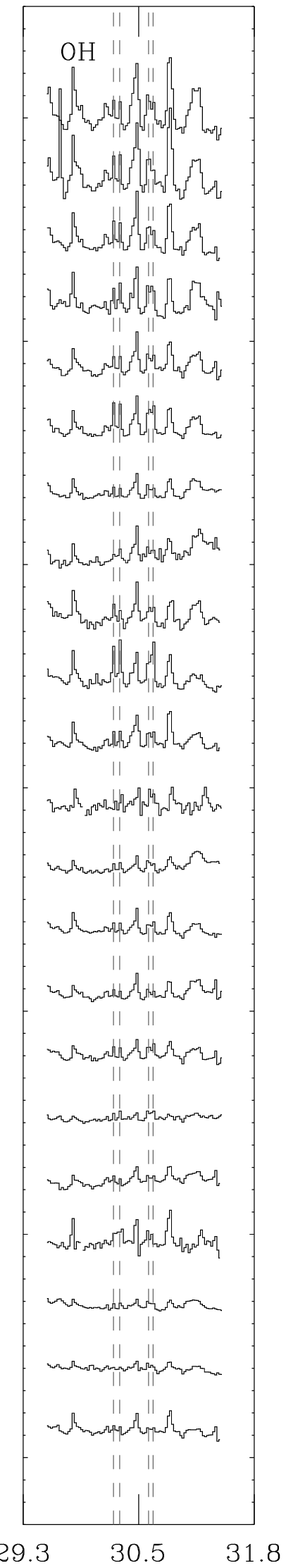}
\includegraphics[width=3.5cm]{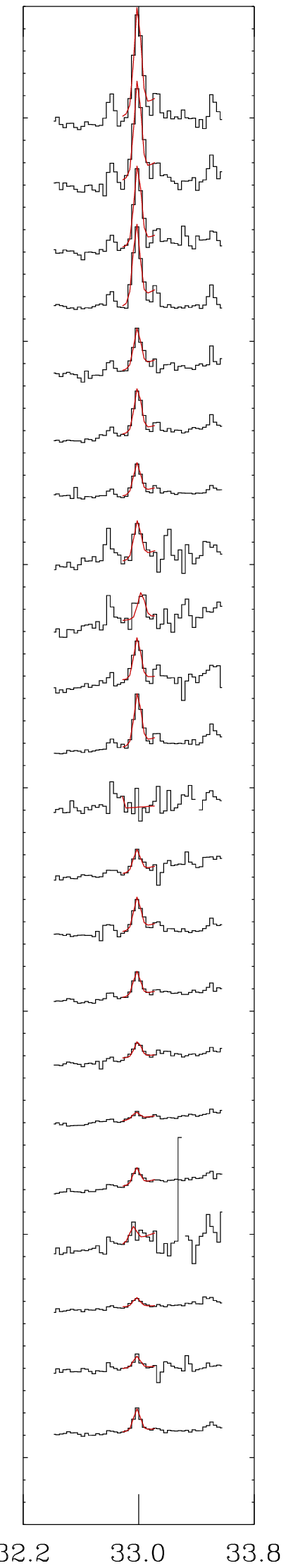}
\caption[]{As for Figure \ref{detectSH}, but for Spitzer-LH data, including the 21.76, 29.85 and 33.0\,$\mu$m H$_2$O complexes, as well as
a region centered on the OH doublets around 30.5\,$\mu$m.}
\label{detectLH}
\end{figure*}

\begin{figure*}
\centering
\includegraphics[width=6cm]{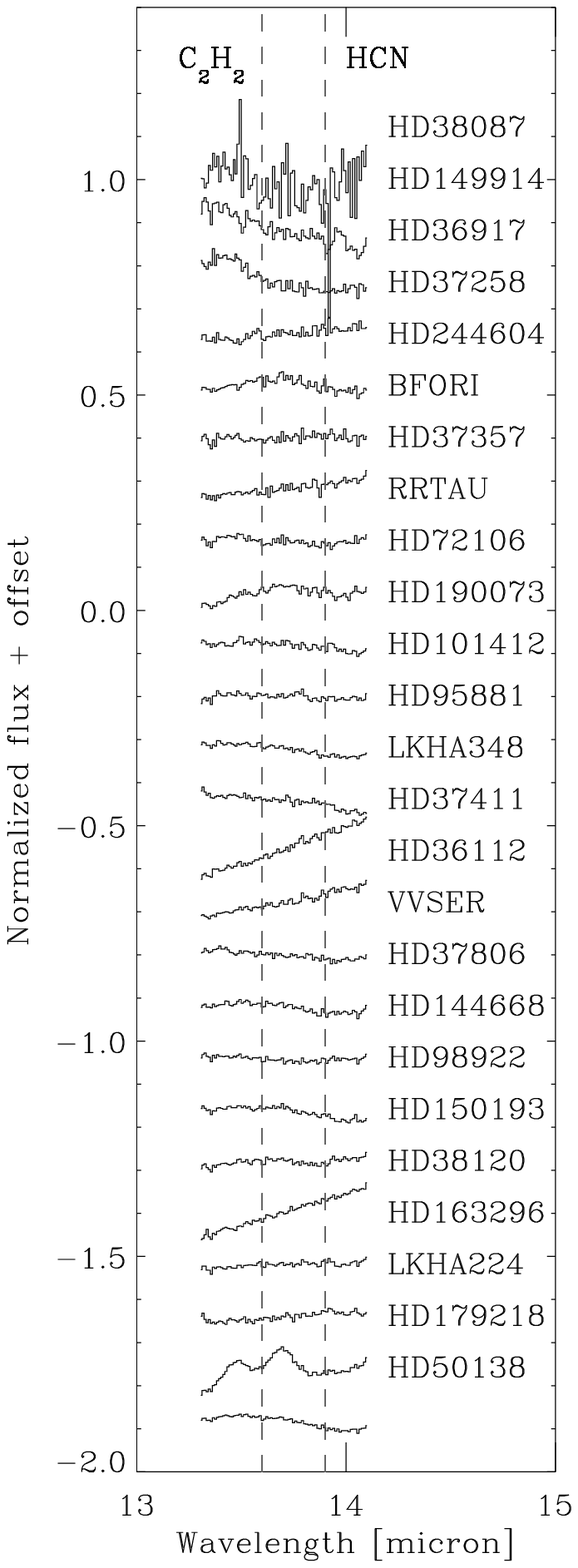}
\includegraphics[width=3cm]{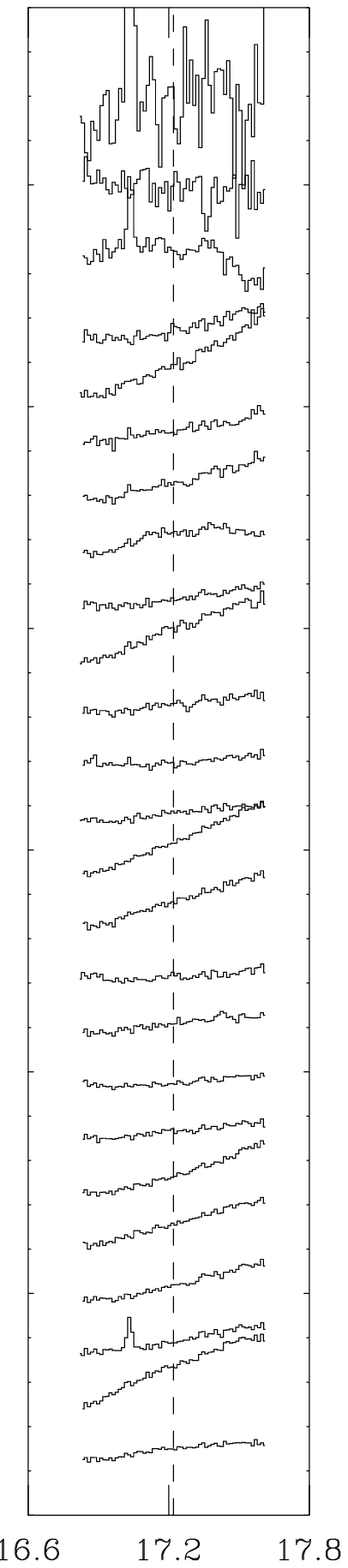}
\includegraphics[width=6cm]{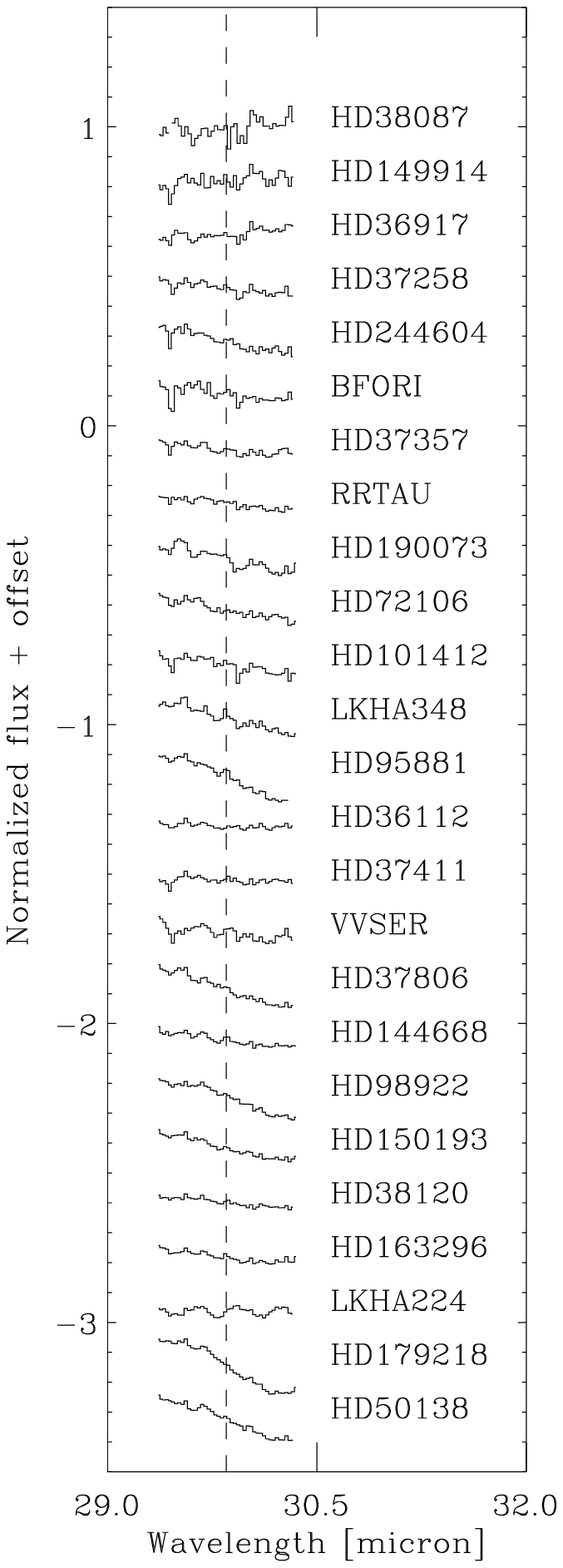}
\caption[]{Selected Spitzer-IRS spectral regions from the full HAe/Be
  disk sample for comparison with Figures \ref{detectSH} and
  \ref{detectLH}. The spectra are ordered according to
  signal-to-noise, with the lowest signal-to-noise at the top. The dashed lines indicate the locations of the tracer water line complexes, except where otherwise noted.}
\label{nondetect}
\end{figure*}

\section{Identification of molecular tracers}

\subsection{Detection criteria}

Due to the large number of lines, almost all of which are blended with other lines of 
similar strength, the analysis of even a single Spitzer spectrum can be a daunting task. 
Further, for sources of somewhat lower quality (signal-to-noise $\sim$ 100), water emission 
close to the detection limit may be difficult to distinguish from noise and residual 
fringe patterns. Therefore, for the purposes of: 1) separating disks where water emission 
is clearly detected from non-detections and borderline cases, and 2) defining a simple 
excitation temperature parameter, a few strong lines were selected that could 
serve as tracers. Such tracer line complexes were selected among features that
are relatively isolated in the spectrum, allowing a continuum to be fitted.  Experiments were made in which a generic LTE model
for water (see paper II) was correlated across a wide wavelength range of the spectrum. However, it was found that this did not 
produce results much different from the approach of using a few line complexes.  
The water line complexes at 15.17 and 17.22\,$\mu$m are more isolated than most, and 
while they are both blends of several lines, the excitation temperature of the transitions 
contributing to each complex fall within a narrow range. The 15.17\,$\mu$m complex traces 
rotational $J$ quantum numbers of 8-10 with excitation energies of 2300-2700\,K, while the 
17.22\,$\mu$m complex traces $J$s of 9-12, corresponding to slightly higher energies of 
2300-3000\,K. Thus, these two complexes trace roughly the same gas, and they are expected to correlate.
 
 A detection of water emission is defined as a detection of both 15.17 and 17.22\,$\mu$m 
complexes at the 3.5$\sigma$ level.
In a few cases, the 15.17\,$\mu$m line was not formally detected, but
other water lines were strong enough to still warrant a clear
detection. Conversely, in some cases lines were formally detected, suggesting the presence of water emission that a visual inspection could not
confirm. For instance, a lack of other water lines in the same spectral range would lead to the source being flagged as 
a non-detection. Due to the nature of the data set, such subjective analysis of a few borderline cases is
unfortunately inevitable. We aimed to be conservative to ensure that there are no false positives in the sample, at the expense
of rejecting a few real detections. Hence, all detection rates can be considered lower limits. Cases that 
were subjectively scrutinized but where we could not unambigously confirm detections of, in particular, water are
flagged in Tables \ref{linefluxes} and \ref{linefluxes_hae}. Selected regions, including those containing the water
tracer lines of the Spitzer-IRS spectra of T Tauri disks for which water is detected are 
shown in Figures \ref{detectSH} and \ref{detectLH}, for the SH and LH
modules, respectively. For comparison, spectral regions including the
water tracer lines as well as the HCN and C$_2$H$_2$ Q branches for the Herbig
Ae/Be disk sample are shown in Figure \ref{nondetect}, although these
spectra are all non-detections. 
In practice, the detection rate does not depend on the exact choice of tracer line complexes, 
except in a few borderline cases. Detections of OH are based on the doublets around 23.0, 
27.6 and 30.5\,$\mu$m, spanning upper level energies of $\sim$2400-4000\,K.  The organics HCN/C$_2$H$_2$ and CO$_2$ are detected via their characteristic 
Q branches. For the latter species, it is noted that there may be a bias against 
their unambiguous detection in sources showing strong water emission, due to blending and 
confusion. In addition, these species are identified based on a single blended feature making it difficult to 
rule out the occasional spurious detection, especially if systematic errors or residual 
fringing exceed calculated errors.  A few such spurious detections were eliminated with 
visual inspection.

The general low contrast of the molecular lines at R=600 relative to the 
strong continuum also affects the detection rate. Specifically, the fringe and flat 
field residuals, in addition to the photon statistics, scale with the continuum, so even for 
very high theoretical signal-to-noise ratios, there is a limit to how low the line-to-continuum 
ratio can be to allow the detection of lines. For our data, this ``fidelity'' limit, i.e. the 
line-to-continuum contrast where the data systematics become larger than the pure
photon statistics, while difficult to quantify, seems to be better than 0.5\%, with several 3.5$\sigma$ detections at the 1-2\% level.
  
\begin{figure}
\centering
\includegraphics[width=8cm]{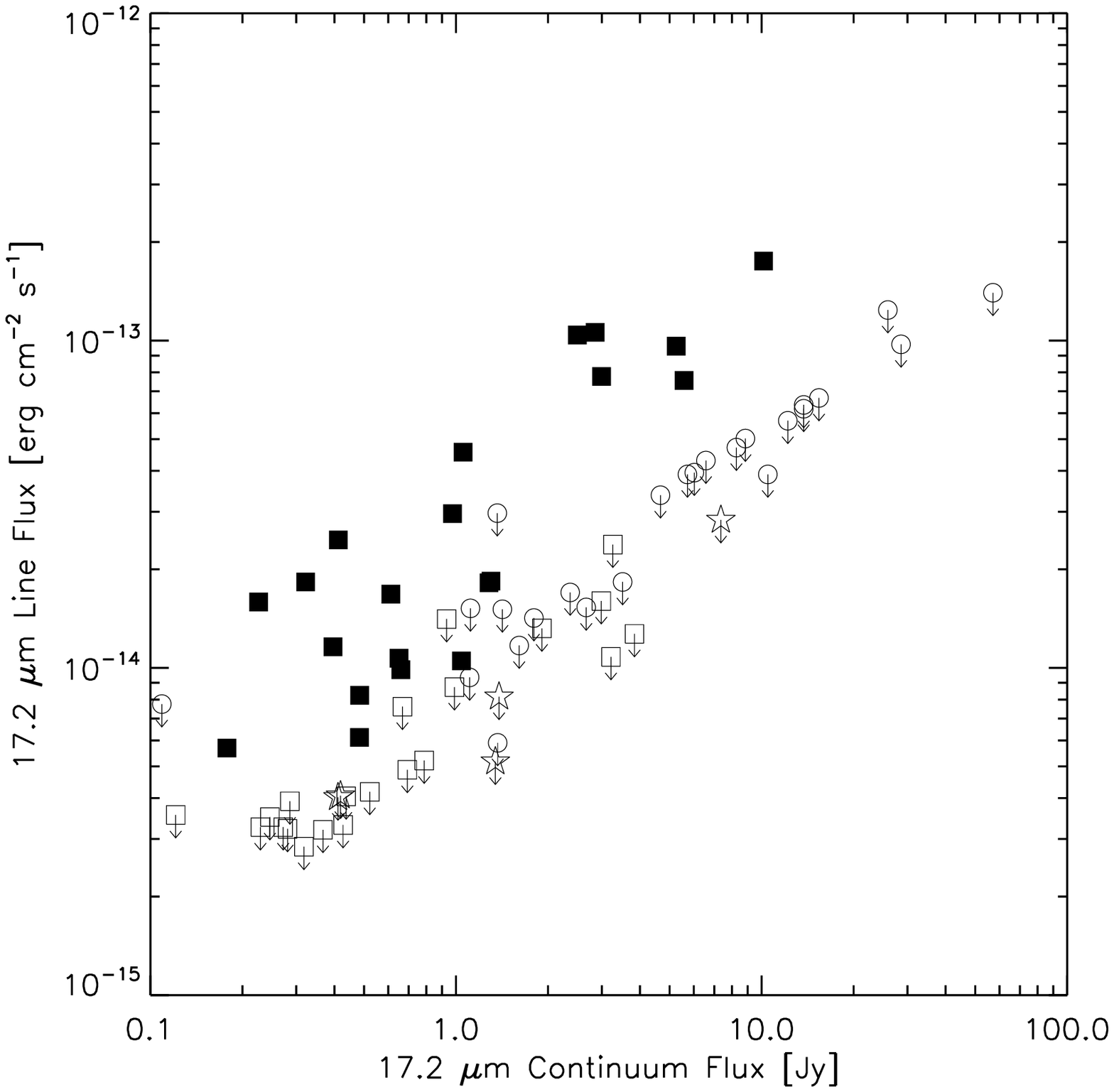}
\caption[]{Relationship between the 17.22\,$\mu$m integrated line flux and the continuum level. 
The filled symbols show detections, while empty symbols with arrows indicate upper limits. 
Squares, stars and circles represent T Tauri stars, transitional objects and Herbig Ae/Be stars, respectively.}
\label{fidelity}
\end{figure}

Figure \ref{fidelity} shows the relation between the continuum level and the flux level of the water tracer. 
The ability to detect a line of constant flux diminishes with increasing flux level, indicating
the ``fidelity'' limit. On the other hand, a significant number of line detections are made in
T Tauri disks with apparent brightnesses higher than those of many Herbig Ae/Be disks. 

\begin{figure}
\centering
\includegraphics[width=8cm]{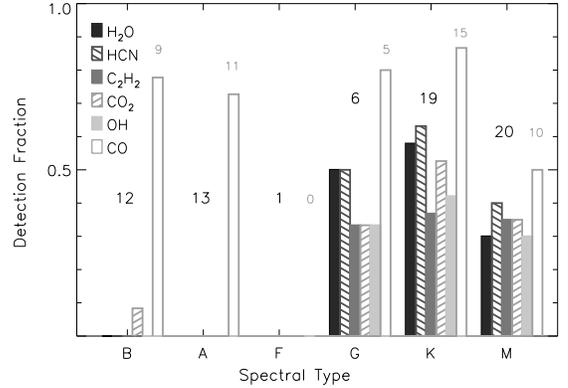}
\caption[]{Detection rate as a function of spectral type for the strongest infrared molecular 
tracers, with the number of each spectral type shown above. The detection rate drops dramatically 
for spectral types earlier than G/F, except for CO. }
\label{detect_hist}
\end{figure}

Using these criteria, water emission is detected in 22 disks, HCN in 25 disks, 
C$_2$H$_2$ in 17 disks, CO$_2$ in 20 disks and OH in 18 disks, as summarized in Figure \ref{detect_hist} and Table \ref{linefluxes}. 
Figure \ref{detect_hist} also shows the detection rates of CO in the rovibrational fundamental band at 4.7\,$\mu$m. The CO observations
are described in Paper II. 

\subsection{H$_2$O excitation tracers}

\begin{figure}
\centering
\includegraphics[width=8cm]{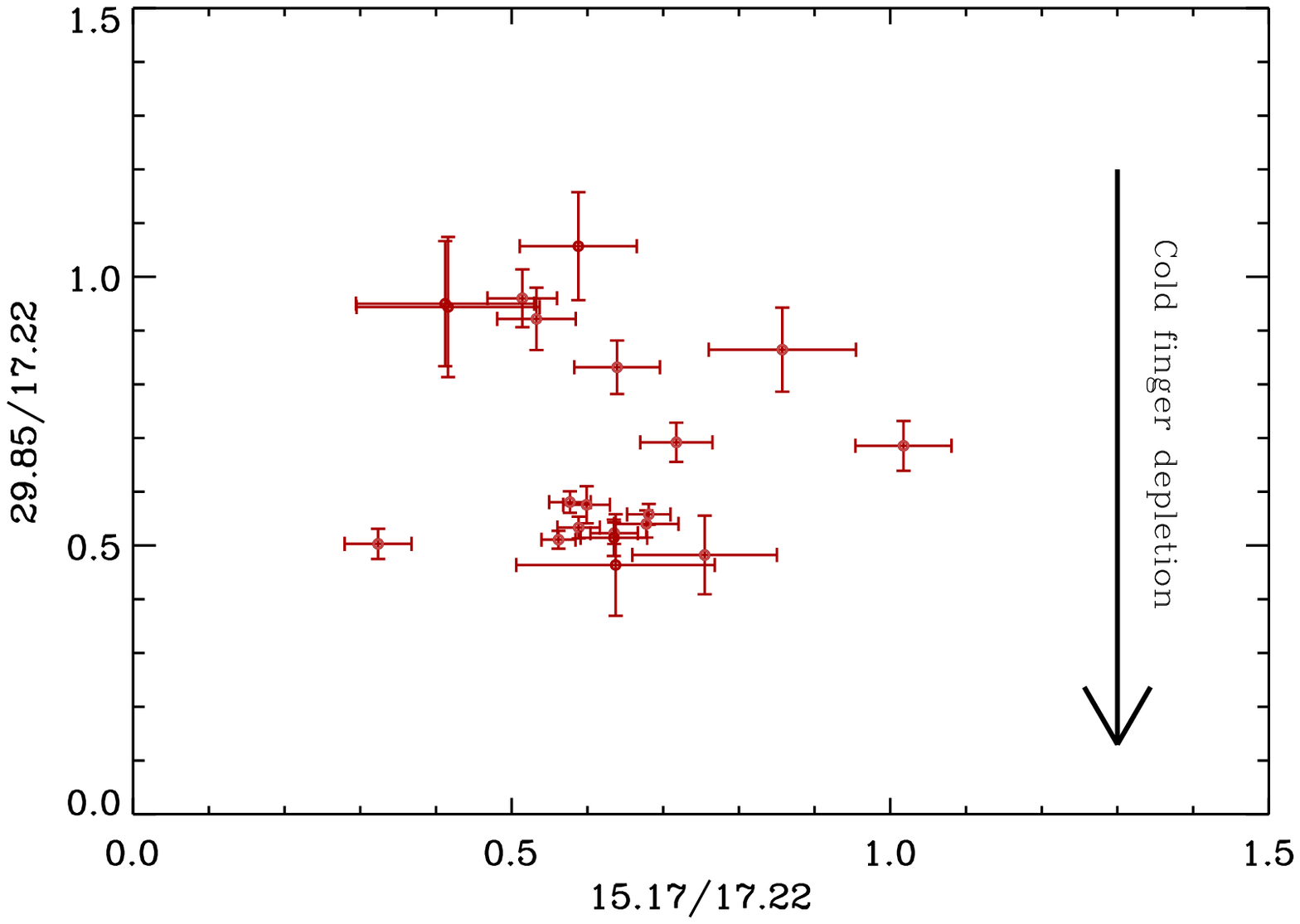}
\caption[]{3-line plot of the observed sample using two high excitation line complexes 
(15.17 and 17.22\,$\mu$m) and one low excitation transition (29.85\,$\mu$m). 
The high excitation lines probe the innermost radii of the disk with temperatures of $T=1000-2000\,$K, 
while the lower excitation line probes larger radii with $T=300-500\,$K. The lower part of the 
plot signifies low water abundances at larger radii, while the upper part of the plot indicates a 
constant surface abundance.}
\label{3line}
\end{figure}

The 15.17 and 17.22\,$\mu$m water line complexes chosen to act as signposts for molecular emission 
trace temperatures in the middle of the range of the 10-36\,$\mu$m water lines. However, the spectra 
also include line complexes from strong transitions with much lower excitation temperatures. 
Ratios of line fluxes between the high and low excitation line complexes are an important diagnostic 
of the nature of the molecular emission; indeed one of the first questions that we ask, based on the 
present sample, is whether the excitation conditions are similar for all disks, or whether there are 
significant differences. As discussed in \cite{Meijerink09} for the particular case of mid-infrared 
water lines, transitions with lower excitation temperature trace larger radii of the disk. Hence, 
line ratios can be affected by both excitation conditions and the radial abundance distribution of 
the species in question.   
In any case, it will be necessary to identify lines with the widest possible baseline in excitation 
energies. A good low energy tracer is the ortho $7_{2\,5}\rightarrow 6_{1\,6}$ line at 29.85\,$\mu$m 
with an excitation energy of 1100\,K (see Figure \ref{H2O_rotladder}). 
Spectra with higher resolving power will be able to separate lines tracing levels at least 
down to $J=5; K_a-K_c=3$ and $E\sim 900\,$K, but these are too blended with higher excitation 
lines at the Spitzer resolution to be clean tracers of specific conditions. Very high excitation 
tracers include lines at 13.3\,$\mu$m (J=15-16, $E=4400-5200\,$K) and 29.5 $\mu$m 
(J=16-18, $E=4600-4800\,$K). 

Table \ref{linefluxes} summarizes the integrated fluxes of the 15.17, 17.22 and 29.85\,$\mu$m line 
complexes. The line fluxes were calculated by fitting a Gaussian superposed on a linear continuum
to the data. The selected line complexes appear to be spectrally unresolved, so the centers and widths of the Gaussian were kept constant for each line complex, corresponding
to the theoretical center of the line complex and the nominal spectral resolving power of Spitzer-IRS.
The wavelength range of the fit (determining where the continuum is constrained) is indicated by the models shown in Figures \ref{detectSH} and \ref{detectLH}.

Figure \ref{3line} shows the distribution of detected water sources in a 3-line plot, matching 
the high excitation 15.17 and 17.22 \,$\mu$m complexes with the low excitation 29.85\,$\mu$m complex. 
Since the low excitation line probes the disk surface at larger radii than the high excitation 
lines, the location of a source in this diagram could provide an indication of the radial water abundance 
structure of the disk surface: the smaller the 29.85/17.22 ratio, the sharper the cutoff of water 
emission beyond a certain radius. This was discussed in \cite{Meijerink09}, who
presented generic non-LTE models of emission spectra, one for a constant abundance of water, and
one in which the surface water abundance was lowered by 6 orders of magnitude beyond a radius of 0.7\,AU, corresponding
to the location of the mid-plane snow-line. 
The arrow in Figure \ref{3line} indicates the approximate range and direction of increasing ``cold finger depletion'' of the surface water abundance
beyond the mid-plane snowline.
It is seen that essentially all disks are located between 
these two extreme cases, with only a few disks exhibiting line ratios that are consistent 
with a high abundance of water throughout the disk surface.  \cite{Meijerink09} suggests a model, 
based on a few H$_2$O Spitzer spectra from \cite{Salyk08} and \cite{Carr08}, in which the disk 
surface water vapor may be strongly depleted at radii larger than that of the midplane snow line 
due to vertical transport of water. Purely chemical effects may also contribute to generate 
structure in the radial surface abundance structure. In any case, the data, as shown in Figure \ref{3line},
indicate that line ratios cluster for many sources, but significant variation does exist in the sample,  with one possible
interpretation being that this is due to structure of the radial emission profile of water.

\begin{deluxetable*}{llllllllll}
\tabletypesize{\scriptsize}
\tablecaption{Line fluxes and molecular detections from the T Tauri star sample}
\tablewidth{0pt}
\tablehead{
\colhead{Source name}   & \colhead{15.17\,$\mu$m\tablenotemark{a}} & \colhead{17.22\,$\mu$m} & \colhead{29.85\,$\mu$m}  & 
\colhead{$L_{\rm H_2O}$\tablenotemark{b}} & \colhead{H$_2$O\tablenotemark{c}} & \colhead{OH}  & \colhead{HCN} & \colhead{C$_2$H$_2$} & \colhead{CO$_2$} \\
                       & $10^{-14}\,\rm erg\,cm^{-2}\,s^{-1}$ &$10^{-14}\,\rm erg\,cm^{-2}\,s^{-1}$ &$10^{-14}\,\rm erg\,cm^{-2}\,s^{-1}$ &$10^{-3} L_{\odot}$& &&&& 
}
\startdata
LkHa 270   & $< 0.36$ & $< 0.34$ & $ 0.82\pm 0.07 $		 & $<1.2$&(0)	  &0	 &0   &0     &1       \\ 
LkHa 271   & $< 0.15$ & $< 0.33$ & $< 0.68 $	                 & $<1.0$&0   &0     &0     &0     &0 \\ 
LkHa 326   & $ 0.47\pm 0.05$ & $ 0.62\pm 0.05$ & $ 0.30\pm 0.04$ & 1.3&1     &1     &1     &0	      &1       \\
LkHa 327   & $ 1.00\pm 0.27$ & $ 2.43\pm 0.25$ & $ 2.31\pm 0.15$ & 5.9&1      &1     &0       &0     &1       \\
LkHa 330   & $< 0.23$ & $< 0.29$ & $< 0.98$			 & $<1.3$&0 &0     &0  & 0 &0\\
LkCa 8     & $< 0.17$ & $ 0.27\pm 0.05$ & $ 0.71\pm 0.06$	 & $<0.3$&(0)	  &0	 &0   &0     &0 	 \\  
IQ Tau     & $ 0.57\pm 0.10$ & $ 0.90\pm 0.09$ & $< 0.51$	 & 0.8&1     &0     &1     &1	      &0	 \\  
V710 Tau   & $< 0.17$ & $ <0.32 $ & $< 0.69$	                 &  $<0.4$& 0   &0    &1     &1     &1 \\ 
AA Tau     & $ 0.48\pm 0.06$ & $ 1.48\pm 0.06$ & $ 0.74\pm 0.03$ & 0.9&1     &1     &1     &1	      &1	\\ 
CoKu Tau/4 & $< 0.29$ & $< 0.33$ & $< 0.39$			 & $<0.3$&0   &0     &0     &0     &0 \\ 
DN Tau     & $< 0.27$ & $< 0.27 $ & $< 0.59 $	                 & $<0.3$ &0	 &0   &0     &0     &0         \\ 
FX Tau     & $< 0.34$ & $< 0.43 $ & $< 0.75 $	                 &  $<0.4$&0	 &0   &1     &0     &0  	\\ 
DR Tau     & $ 4.53\pm 0.19$ & $ 7.13\pm 0.19$ & $ 3.73\pm 0.10$ & 4.4&1     &1     &1     &1	  &1	      \\ 
SX Cha     & $ 0.78\pm 0.06$ & $ 1.21\pm 0.06$ & $ 1.01\pm 0.04$ & 1.3&1     &1     &1     &0	      &0	\\
SY Cha     & $< 0.09$ & $ 0.12\pm 0.03$ & $ 0.43\pm 0.02$	 & $<0.2$ &(0)     &0	  &1	      &1     &1 	\\ 
TW Cha     & $ 0.61\pm 0.02$ & $ 1.04\pm 0.03$ & $ 0.55\pm 0.02$ & 1.2&1     &1     &1     &1	      &0      \\  
VW Cha     & $ 1.99\pm 0.07$ & $ 3.54\pm 0.08$ & $ 1.81\pm 0.04$ & 3.6&1     &1     &1     &1	      &1       \\  
VZ Cha     & $ 0.94\pm 0.05$ & $ 1.39\pm 0.05$ & $ 0.75\pm 0.02$ & 1.4&1     &1     &1     &1	      &0	\\  
WX Cha     & $ 1.09\pm 0.04$ & $ 1.61\pm 0.04$ & $ 0.90\pm 0.02$ & 1.6&1     &1     &1     &1	      &1	\\ 
XX Cha     & $ 0.40\pm 0.03$ & $ 0.46\pm 0.03$ & $ 0.40\pm 0.02$ & 0.5&1     &1     &1     &1	      &1	\\  
T Cha      & $< 0.32$ & $< 0.31$ & $< 0.92$			 & $<0.08$&0   &0     &0       &0     &0 	  \\ 
Sz 50      & $< 0.12$ & $< 0.13$ & $< 0.18$			 & $<0.2$&0  &0      &0     &0     &1	 \\  
HD 135344B & $< 0.32$ & $< 0.35$ & $< 2.11$		         & $<0.2$&0   &0     &0       &0     &0\\ 
HT Lup     & $< 1.75$ & $< 1.72$ & $ 1.17\pm 0.15$		 & $<1.1$&(0)	  &0	 &0   &0     &1 	\\   
GW Lup     & $< 0.17$ & $< 0.18$ & $< 0.34$		         & $<0.3$&0	  &0	 &0   &0     &1 	\\ 
GQ Lup     & $ 0.71\pm 0.06$ & $ 1.39\pm 0.06$ & $ 1.33\pm 0.04$ &1.3 &1     &1     &1     &0	&0     \\ 
IM Lup     & $< 0.24$ & $< 0.24$ & $< 0.30$			 & $<0.2$&0	&0     &0     &0     &1 	 \\ 
HD 142527  & $< 1.96$ & $< 2.22$ & $< 5.79$			 & $<0.5$& 0  &   0   &  0    & 0     &0 \\
RU Lup     & $ 2.50\pm 0.13$ & $ 3.49\pm 0.14$ & $ 2.41\pm 0.08$ & 2.8&1     &0     &1     &0	      &1	\\ 
RY Lup     & $< 0.65$ & $< 0.74$ & $< 0.97$			 & $<0.7$&0   &0      &0     &0     &0  	\\ 
EX Lup     & $< 0.37$ & $ 1.28\pm 0.12$ & $ 4.05\pm 0.07$	 &2.4 &1     &1     &0     &0	      &0	\\ 
AS 205     & $11.55\pm 0.65$ & $18.19\pm 0.74$ & $ 9.35\pm 0.48$ &8.9 &1     &1     &1     &1	      &1       \\ 
Haro 1-1   & $< 0.23$ & $< 0.22$ & $< 0.67$			 & $<0.3$&0   &0     &0     &0     &0 \\  
Haro 1-4   & $< 0.38$ & $< 0.47$ & $< 0.75$		         & $<0.4$&0   &0     &1     &1     &0 \\
VSSG1      & $ 1.54\pm 0.18$ & $ 2.63\pm 0.17$ & $ 2.78\pm 0.19$ & 1.9&1     &0     &1     &1	      &1      \\
DoAr 24E   & $ 2.65\pm 0.11$ & $ 4.42\pm 0.12$ & $ 2.54\pm 0.13$ & 2.5&1     &1     &1     &1	      &1  \\
DoAr 25    & $< 0.21 $ & $< 0.51 $ & $< 0.53$	                 & $<0.3$&0   &0     &1     &1     &0 \\ 
SR 21      & $< 1.04$ & $< 1.29$ & $< 3.90 $		 & $<0.2$&0   &0     &0     &0     &0	\\     
SR 9       & $< 1.05 $ & $< 0.70 $ & $ 1.32\pm 0.08$ & $<0.5$    &(0)   &0     &0     &0     &0	 \\	 
V853 Oph   & $< 0.28$ & $ 0.55\pm 0.08$ & $ 1.14\pm 0.05$	 & 0.6&1     &1     &1     &1	      &0 \\ 
ROX 42C    & $< 0.38$ & $< 0.37$ & $< 0.68$			 & $<0.3$&0   &0     &0     &0     &0  \\ 
ROX 43A    & $< 0.85$ & $< 0.92$ & $< 1.36$		         & $<0.7$&0   &0     &0     &0     &0  \\ 
Haro 1-16  & $ 0.83\pm 0.07$ & $ 1.56\pm 0.08$ & $ 1.44\pm 0.06$ & 1.0&1     &1     &1     &0	      &0 \\
Haro 1-17  & $< 0.14$ & $< 0.14$ & $< 0.35$			 & $<0.2$&0	&0     &0     &0     &0 \\ 
RNO 90     & $ 5.83\pm 0.24$ & $10.10\pm 0.25$ & $ 5.86\pm 0.14$ & 5.8&1     &1     &1     &1	  &1	      \\   
Wa Oph/6   & $ 1.57\pm 0.07$ & $ 1.54\pm 0.07$ & $ 1.06\pm 0.05$ & 0.8&1     &1     &1     &0	      &1\\
V1121Oph   & $ 1.12\pm 0.30$ & $ 2.69\pm 0.32$ & $ 2.54\pm 0.17$ & 2.3&1     &0     &0     &0	      &0\\
EC 82      & $< 0.31$ & $< 0.35$ & $< 0.65$		 & $<4.9$&0   &0     &0     &0     &0  \\     
\enddata

\tablenotetext{a}{All upper limits are 3.5\,$\sigma$; errors are 1\,$\sigma$.}
\tablenotetext{b}{The integrated line flux due to H$_2$O in the Spitzer wavelength range. Water has a multitude of lines outside the
observed spectral range, including the strong rovibrational band around 6\,$\mu$m. Extrapolating to all water lines may significantly
increase the total cooling rate.}
\tablenotetext{c}{(0)  indicates a tentative detection of water based on an inspection by eye. Some times, but not always, these borderline detections are 
accompanied by formal detections of specific line complexes. Conversely, in a few cases, formal detections were made that could not be confirmed by a 
visual inspection, likely due to interference from data artifacts.}

\label{linefluxes}
\end{deluxetable*}

\begin{deluxetable*}{llllllllll}

\tablecaption{Line fluxes and molecular detections from the Herbig Ae/Be star sample}
\tablehead{
\colhead{Source name}   & \colhead{15.17\,$\mu$m\tablenotemark{a}} &\colhead{17.22\,$\mu$m}& \colhead{29.85\,$\mu$m} &  \colhead{$L_{\rm H_2O}$\tablenotemark{b}} & \colhead{H$_2$O\tablenotemark{b}} & \colhead{OH}  & \colhead{HCN} & \colhead{C$_2$H$_2$} & \colhead{CO$_2$} \\
                       & \colhead{$10^{-14}\,\rm erg\,cm^{-2}\,s^{-1}$} &\colhead{$10^{-14}\,\rm erg\,cm^{-2}\,s^{-1}$} &\colhead{$10^{-14}\,\rm erg\,cm^{-2}\,s^{-1}$} &$10^{-3} L_{\odot}$& &&&& 
}
\startdata
HD 36112      & $< 1.58$ & $< 1.93$ & $< 3.84$		     &$<5.8$& (0)  &   0	&  0	& 0	&0 \\
HD 244604     & $< 0.86$ & $< 0.83$ & $< 0.91$		     &$<5.9$& 0  &   0	&  0	& 0	&0 \\
HD 36917      & $< 0.94$ & $< 1.01$ & $< 1.44$		     &$<9.7$& 0  &   0	&  0	& 0	&0 \\
HD 37258      & $< 0.80$ & $< 0.89$ & $< 0.70$	             &$<6.2$& (0)  &   0	&  0	& 0	&0 \\
BF Ori        & $< 0.62$ & $< 0.59$ & $< 0.72$		     &$<3.9$& 0  &   0	&  0	& 0	&0 \\
HD 37357      & $< 0.78$ & $< 0.79$ & $< 1.11$		     &$<11.3$& 0  &   0	&  0	& 0	&0 \\
HD 37411      & $< 0.38$ & $< 0.38$ & $< 1.05$		     &$<8.0$& 0  &   0	&  0	& 0	&0 \\
RR Tau        & $< 0.69$ & $< 0.68$ & $< 0.93$		     &$<172$& (0)  &   0	&  0	& 0	&0 \\
HD 37806      & $< 2.01$ & $< 2.09$ & $< 2.45$		     &$<38$& (0)  &   0   &  0    & 0     &0 \\ 
HD 38087      & $< 0.22$ & $< 0.27$ & $< 0.87$		     &$<5.7$& 0  &   0	&  0	& 0	&0 \\
HD 38120      & $< 2.15$ & $< 2.62$ & $< 4.39$		     &$<58$& 0  &   0   &  0    & 0     &0 \\
HD 50138      & $< 4.57$ & $< 4.99$ & $ 7.81\pm 0.89$	     &$<84$& (0)  &   0   &  0    & 0     &0 \\
HD 72106      & $< 0.86$ & $< 0.88$ & $< 1.16$		     &$<4.0$& (0)  &   0	&  0	& 0	&0 \\
HD 95881      & $< 1.78$ & $< 1.75$ & $< 2.37 $	             &$<1.4$& 0  &   0	&  0	& 0	&0 \\
HD 98922      & $< 4.66$ & $< 4.83$ & $< 3.96$		     &$<346$& (0)  &   0	&  0	& 0	&0 \\
HD 101412     & $< 0.86$ & $< 0.82$ & $< 0.82$		     &$<1.1$& (0)  &   0	&  0	& 0	&1 \\
HD 144668     & $< 2.42$ & $< 2.45$ & $ 4.25\pm 0.53$	     &$<8.7$& (0)  &   0	&  0	& 0	&0 \\
HD 149914     & $< 0.21$ & $< 0.16$ & $< 0.36$		     &$<0.16$& 0  &   0   &  0	 & 0	 &0 \\
HD 150193     & $< 2.35$ & $< 2.61$ & $< 4.04$		     &$<5.8$& (0)  &   0	&  0	& 0	&0 \\
VV Ser        & $< 0.94$ & $< 1.15$ & $ 0.99\pm 0.14$        &$<11.7$& (0)  &   (0)	&  0	& 0	&0 \\
LkHa 348      & $< 1.89$ & $< 1.78$ & $< 1.98$	             &$<13.3$& 0  &   0	&  0	& 0	&0 \\
HD 163293     & $< 2.42$ & $< 2.73$ & $ 4.34\pm 0.39$	     &$<4.3$& (0)  &   0	&  0	& 0	&0 \\
HD 179218     & $< 3.14$ & $< 3.76$ & $< 6.26$		     &$<54$& 0  &   0   &  0    & 0     &0 \\
HD 190073     & $< 1.64$ & $< 1.66$ & $< 1.75$		     &$<65$& 0  &   0	&  0	& 0	&0 \\
LkHa 224      & $< 1.52$ & $< 1.57$ & $< 2.04$		     &$<219$& 0  &   0	&  0	& 0	&0 \\

\enddata
\tablenotetext{a}{All upper limits are 3.5\,$\sigma$; errors are 1\,$\sigma$.}
\tablenotetext{b}{(0) indicates a tentative detection of water based on an inspection by eye. Some times, but not always, these borderline detections are 
accompanied by formal detections of specific line complexes. Conversely, in a few cases, formal detections were made that could not be confirmed by a 
visual inspection, likely due to interference from data artifacts. }

\label{linefluxes_hae}
\end{deluxetable*}

\section{Discussion}

\subsection{Cooling balance}

The temperature structure of protoplanetary disks is determined by a detailed balance of heating and cooling processes. 
In the surface layers, collisional exchange between the gas and the dust is not sufficient to maintain a temperature equilibrium
between the dust and the gas, leading to elevated gas temperatures. The degree to which the gas is superheated depends
sensitively on the efficiency of line cooling. As a consequence, the finding that the mid-infrared wavelength range in typical protoplanetary disks is blanketed 
in molecular lines will have important consequences for models of disk structures. 

Current disk models find water cooling rates of a few $\times 10^{-5}\,\rm L_{\odot}$ and individual cooling line luminosities from atomic species, as well as rotational lines of CO and H$_2$, are $10^{-8} - 10^{-5}\,\rm L_{\odot}$ \citep{Gorti08}. However, we measure integrated water line luminosities in the 10-36\,$\mu$m region that are 1-2 orders of magnitudes higher. 
Table \ref{linefluxes} presents the 10-37\,$\mu$m integrated H$_2$O
line luminosities for the disks in which H$_2$O has been detected. 
The line luminosities were determined by scaling the fiducial model of
\cite{Meijerink09} to the SH and LH spectra, where different scaling
values were allowed for the two spectral ranges. Typical values
resulting from this range from $<5\times 10^{-4}$ to almost
$10^{-2}\,\rm L_{\odot}$. Extrapolating to include lines outside the Spitzer 
range may increase these numbers by a factor 2 or more. 
Consequently, the detection of the mid-infrared molecular spectra from 
protoplanetary disks increases the total disk-averaged cooling rates 
by 1-2 orders of magnitude, relative to recent models, for gas temperatures 
below $\lesssim 2000\,$K; the infrared cooling budget is dominated by
water in T Tauri stars. It should be stressed, however, that 
most of the water lines in the Spitzer range are likely formed in the
innermost ($<$ a few AU) regions of the disk, the outer regions of which 
may be less affected by water as an exceptionally strong coolant. 
An immediate consequence of more efficient cooling seems to be that the 
surface layers of the disks will have a larger column of relatively cooler gas, 
facilitating the survival, and observability, of a rich chemistry in the disk surface.

\subsection{Lack of detectable molecular emission from Herbig Ae/Be stars}

\begin{figure}
\centering
\includegraphics[width=8cm]{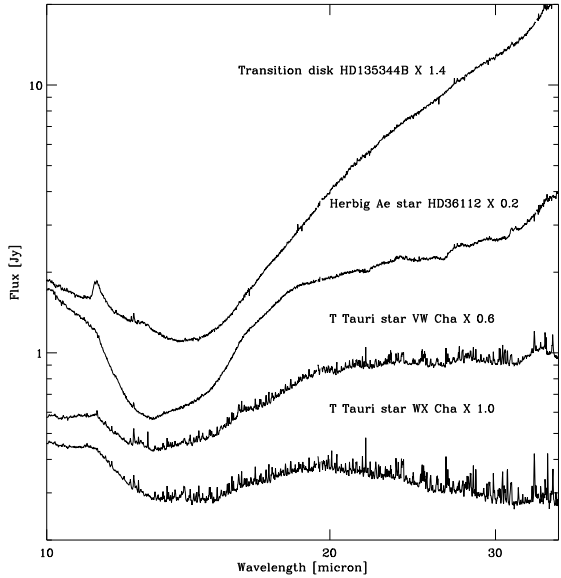}
\caption[]{Comparison of typical spectra from three classes of objects: 
Transitional disks, Herbig Ae/Be stars and classical T Tauri stars. 
The spectra have been scaled for clarity (not shifted), preserving the line-to-continuum ratios.}
\label{ctts_hae}
\end{figure}

\begin{figure}
\centering
\includegraphics[width=8cm]{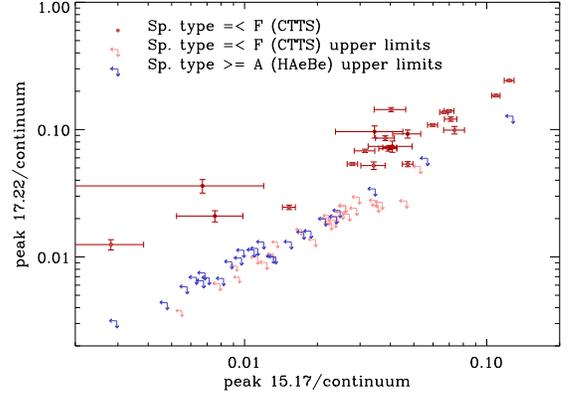}
\caption[]{Distribution of line-to-continuum ratios of the 15.17 and 17.22\,$\mu$m line complexes. 
The sample has been split into T Tauri stars (spectral types later or equal to F) and Herbig Ae stars 
(spectral types earlier than F). }
\label{TTvsHAe}
\end{figure}

Arguably the strongest observation, based on the current sample, is that the detection rate of 
molecular emission from Herbig Ae/Be stars with Spitzer-IRS is apparently very low. The upper limits
for detections of the water tracers are given in Table \ref{linefluxes_hae}. In fact, 
out of 25 Herbig Ae stars, there is not a single detection of water, HCN, C$_2$H$_2$ or OH, given our detection criteria;
CO$_2$, which is detected in HD101412, provides the sole exception. In comparison, the detection rate for disks around 
stars with spectral types later than F is $\sim$40\%.  Could this effect be due to some bias? 
First of all, as Figure \ref{spindex} shows, the sample of Herbig stars are 
somewhat brighter than the T Tauri stars, by up to an order of magnitude. This indicates that 
the signal-to-noise ratios of the Herbig star spectra are unlikely to be significantly
lower than those of the T Tauri stars. Figure \ref{ctts_hae} shows a comparison between the full IRS spectra of a 
Herbig star, a transitional disk and two classical T Tauri stars. Further, in Figure \ref{TTvsHAe} a comparison of 
the line-to-continuum ratios of the H$_2$O line tracers at  15.17 and 17.22\,$\mu$m is given
for the full sample. It is seen than the Herbig stars have 3.5$\sigma$ upper limits on 
their line-to-continuum ratios that are systematically smaller by a factor of 5-10 than the ratios in T Tauri stars where water is detected. 

\begin{figure}
\centering
\includegraphics[width=7cm]{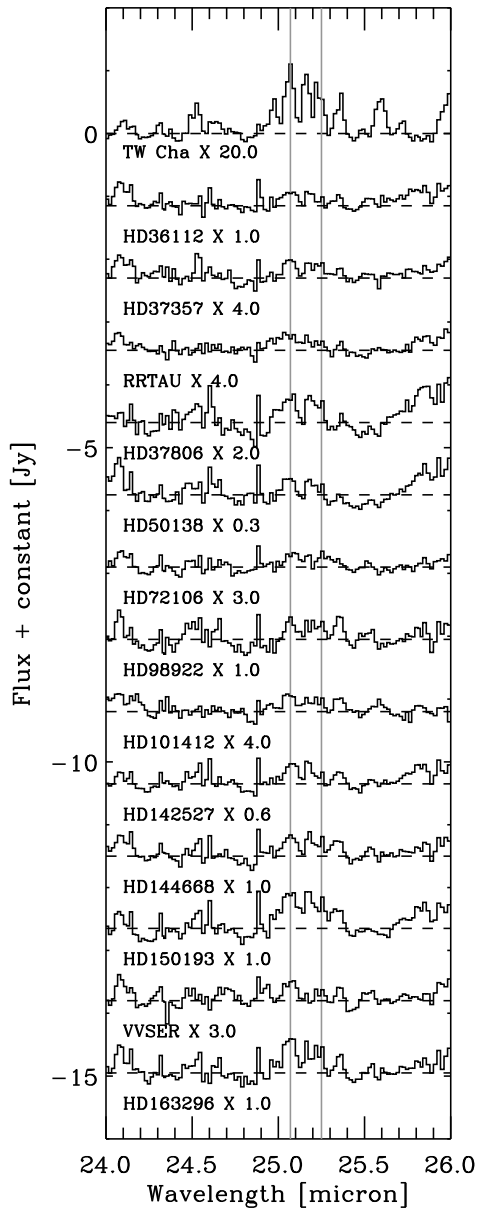}
\caption[]{Tentative detections of line complexes due to a combination of H$_2$O and OH lines in the sample
of Herbig Ae/Be disks, specifically, 
the complex located between 24.9 and 25.5\,$\mu$m. The top spectrum of
TW Cha is displayed for comparison with a source with clear detections.
The two vertical lines indicate the locations of two OH doublets.
Due to the low line-to-continuum of the lines and the possibility of
interference from systematics at the same level, we refrain
from analyzing these further. }
\label{coolwater}
\end{figure}

It is interesting to note that some Herbig Ae/Be disks do show tentative low-level emission features
that may be due to water and OH at longer wavelengths, specifically in the LH module. Such tentative detections
made using the 24.9-25.5\,$\mu$m OH/H$_2$O complex are displayed in Figure \ref{coolwater} and compared
to the water spectrum from TW Cha.  We describe them
as tentative because they do not unambiguously match a water model over a wider range of wavelengths, although that
could be explained by noise or systematics that vary with wavelength or spectral order. The 25\,$\mu$m complex
covers the $\rm X\,\pi_{3/2}\rightarrow X\,\pi_{3/2}\,10.5$ and $\rm X\,\pi_{1/2} \rightarrow X\,\pi_{1/2}\,9.5$ OH lines and water lines spanning excitation energies of 1500-2500\,K, or somewhat cooler than 
those traced by the 15.17 and 17.22\,$\mu$m complexes. If real, these detections still represent line-to-continuum ratios significantly lower than those
of T Tauri disks, but suggest that, with a modest improvement in data quality, molecular emission features should
also be detected in Herbig Ae/Be disks. Given that the tentative detections are at the same level of the current data
systematics, it is not possible to further analyze them. Their presence, however, do suggest that the optically thick H$_2$O lines
at far-infrared wavelengths may be detected in Herbig Ae/Be disks by Herschel.

While the mid-infrared molecular line-to-continuum ratios are much smaller in Herbig Ae/Be stars relative to those
in T Tauri stars, is it possible that the line fluxes are similar, but veiled by the stronger
continuum fluxes of the Herbig stars?  Most disks around Herbig Ae/Be stars, as well most transitional disks, do show strong lines
from CO at the rovibrational band at 4.7\,$\mu$m \citep[][and Paper II]{Blake04, Salyk09}. New ground-based CO observations
of the full sample are discussed in greater detail in Paper II.  

A comparison of line-to-continuum ratios for CO M-band lines from \citet{Najita03} and \citet{Blake04} 
demonstrates lower line-to-continuum contrast, on average, for Herbig Ae/Be vs. T Tauri disks. The reason for this
difference is still unclear, but if the water and other molecules behave similarly, this may explain the general lack
of high contrast molecular line emission from Herbig Ae/Be stars. 
However, the difference for CO is only $\sim$ a factor of 2 --- significantly smaller than the minimum difference 
required for water. 

It is therefore tempting to conclude that the observed lower line-to-continuum ratios in Herbig Ae/Be disks are produced by lower molecular abundances, 
although other alternative explanations are possible. Differences in dust properties could affect line strengths, or the heating rate of the molecular layer may not scale 
linearly with stellar luminosity. Here, we discuss these possibilities.

\subsubsection{Photochemistry} 

In principle, the harsher radiation field from A and B stars effectively photodissociates molecules in unshielded regions of 
the disk surface at 1-2 AU, which could result in lower line fluxes relative to the
disk luminosity. If photochemistry is important, one would expect to see strong chemical changes with spectral type, such as an increasing OH/H$_2$O 
abundance ratio for disks around earlier type stars. Some early indications of this
have already been found in two studies reporting detections of OH, but not H$_2$O, 
for one Herbig Ae/Be star by \cite{Mandell08} and in a transitional
disk \citep{Najita10}. A dependence of inner disk emission on spectral type between late type T Tauri stars and brown dwarfs has also been found for the HCN/C$_2$H$_2$ ratio
by \cite{Pascucci09}, although in this case photochemistry is unlikely to be the cause of the observed difference. 
Photochemistry may also give rise to a dichotomy between molecules (like CO)
that can efficiently self-shield and those that cannot; molecular self
shielding may be a very strong effect in disks, i.e. even stronger than in
molecular clouds because the continuum dust opacity is expected to be small due to a depletion of dust opacity
in the disk surface following dust growth and settling to the
midplane. 

Water is an interesting case for self-shielding that was recently studied for the
case of protoplanetary disks by \cite{Bethell09}.  These authors show
that H$_2$O, as well as OH, can in fact self-shield, although at higher column
densities than CO. Specifically, water self-shields at column
densities of $N_{\rm H_2O}\sim 2\times 10^{17}\,\rm cm^{-2}$, while CO self-shields
at $N_{\rm CO} \sim 5\times 10^{15}\,\rm cm^{-2}$ \citep{vandishoeck88,visser09} in a dust-free
environment, and assuming abundances of $\sim 10^{-4}$ relative to
H. This suggests that CO may survive closer to the star in regions unshielded by dust, 
predicting that low excitation water lines may still be seen from larger radii in Herbig disks, even
if water is absent from warm gas. CO emission is generally
observed in the inner regions of classical Herbig Ae/Be disks
\citep[e.g.][]{Blake04}, including the majority of the disks in the sample
presented here (Paper II). Note that some disks around Herbig Ae/Be stars are also part of the 
class of transitional disks, and some of those appear to have
strongly depleted CO abundances in their inner disks \citep{vanderplas09,Brittain09}, but
that may be related to their evolutionary stage, rather than to an intrinsic property
of Herbig Ae/Be disks.

\cite{Woitke09} modeled the abundance of water vapor
throughout the disk surrounding a Herbig Ae star, including a
detailed treatment of UV-driven photochemistry, and found that the
inner disk abundance is high, if shielded, reaching $10^{-5} - 10^{-4}$ per
hydrogen nucleus, and can maintain abundances of $10^{-6}\,\rm H^{-1}$ in higher, 
unshielded layers. While there are no predicted line fluxes for the Spitzer wavelength range, 
this model does produce strong lines at wavelengths $>70\,\mu$m. Line spectra in the
Spitzer windows were generated by \cite{Pontoppidan09} and \cite{Meijerink09} for 
similar water abundances, but for a lower mass star, which resulted in strong water lines. 
A detailed radiative transfer comparison of mid-infrared molecular lines from the disks around stars 
of varying stellar mass is therefore needed, and current chemical disk models may have
to consider additional effects to explain the observed difference between T Tauri 
and Herbig Ae/Be disks. 

\subsubsection{Relative scaling of line and continuum luminosities}
\label{scaling}
As is seen in Tables \ref{linefluxes} and \ref{linefluxes_hae}, the upper limits on the total H$_2$O line luminosities 
for the Herbig star sample tend to be higher than the line luminosities for the T Tauri stars with detected lines, with a 
few exceptions. Because the mid-IR continuum level is generally proportional to the stellar luminosity, it is possible for the molecular 
spectra to remain undetected in the Herbig Ae/Be stars if the molecular line luminosity is a significantly weaker function of stellar luminosity.
Ultimately, the scaling of line fluxes with the stellar luminosity may be related to how the regions of the disk surface that form the molecular lines are heated. Three principal
sources of disk surface heating have been identified: Heating by the stellar optical/infrared continuum of dust and subsequent collisional
coupling to the gas, photo-electric heating by UV photons \citep{Jonkheid04,Kamp04,Nomura05} 
and ionization heating by X-rays \citep{Glassgold04}. 

The dust continuum is generated by the first of these processes, and
if that dominates the heating of the line-forming layer, the line luminosity can be expected to scale similarly with stellar luminosity, resulting in no significant line-to-continuum dependence 
with spectral type. The strength of the UV radiation field is a strongly increasing function of stellar effective temperature (and for T Tauri stars, of accretion rate) and
would, in the absence of chemical differences, be expected to increase line-to-continuum ratios with increasing stellar effective temperature. 
The X-ray luminosity, on the other hand, is a much weaker function of stellar type, although highly variable, and may, 
in median, only differ by 1-2 orders of magnitude between T Tauri stars and Herbig Ae/Be stars \citep{Feigelson05}.
If the molecular line luminosity is required to be, at most, constant with spectral type, and in the absence of differences 
in chemistry and excitation, X-ray heating of the molecular layer would likely be the best candidate for determining
the line luminosity. Regardless, photo-electric heating becomes weak at column densities in excess of $10^{21}\,\rm cm^{-2}$ and
tends to be associated with layers with low molecular abundances \citep{Dullemond07}. Given that the column
densities of the layers emitting the mid-infrared molecular lines are likely higher $\gtrsim 10^{22}\,\rm cm^{-2}$ \citep{Carr08,Salyk08}, 
the mechanism for gas heating in the uppermost layers may not be important. If photo-electric heating, by X-rays or otherwise, is not the dominant 
heating mechanism in the line-forming regions, dust coupling is the main candidate, leading to a strong dependence of the line luminosity on stellar spectral type. 
In conclusion, we find it unlikely that the molecular line luminosity is a sufficiently weak function of stellar luminosity, in the absence of chemical or other systematic 
differences in the structures of disks around T Tauri relative to those surrounding Herbig Ae/Be stars. 

\subsubsection{Dust properties}
The lines from Herbig stars could be intrinsically weaker due to a relative difference in the 
gas-to-small-dust ratio in the disk atmosphere leading to a reduced column of observable molecules.  
This would require different relative rates of vertical mixing and settling of grains for the two types of disks. Specifically, 
if the disk surfaces of Herbig Ae/Be stars are more turbulent, the water column above the $\tau=1$ surface of dust may not be
sufficient to form strong lines. This scenario was modeled by \cite{Meijerink09}, who indeed found that
the observed line strength in T Tauri disks required high gas-to-dust ratios of $\sim 10^4$. If the gas and dust
in Herbig Ae/Be disks are mixed closer to the canonical gas-to-dust ratio of 100, this would translate to suppressed line strengths. 
The higher ionizing radiation fields around Herbig Ae/Be stars may allow for a better coupling of the disk surface to 
the stellar magnetic field, thus activating the magneto-rotational-instability and driving a larger degree of surface turbulence \citep{Balbus98}. 
A highly turbulent disk would also work to counteract the effect in which water (and therefore oxygen) is depleted from the disk surface due to
freeze-out and settling \citep{Meijerink09}.  

A difficulty with this scenario is that it predicts that lower line-to-continuum ratios should be observed across the infrared range for
all observed molecular lines, including the CO ro-vibrational lines at 4.7\,$\mu$m. Since these lines are generally observed in Herbig Ae/Be disks, with column densities
similar to those observed in T Tauri disks \citep{Najita03,Blake04}, it is not obvious that there is a general difference in gas-to-dust ratio
in T Tauri disks relative to Herbig Ae/Be disks. Also, X-rays, as compared to UV, are more effective in generating ionization in layers deep enough to facilitate
efficient turbulent mixing \citep{Igea99}, but X-rays will not scale as rapidly with stellar type as UV as discussed section \ref{scaling}, arguing against this mechanism.

\subsection{Lack of molecular emission from transitional disks}

Transitional disks are protoplanetary disks with a deficit of infrared emission due to a paucity 
of dust in the inner regions \citep{Strom89}. This class of disks is thought to be
in the process of clearing out their inner regions through various processes, possibly including
planet formation. There are 5 transitional disks in this sample, 
LkHa 330, SR 21, T Cha, HD135344B and CoKu Tau/4. However, none show H$_2$O, HCN, C$_2$H$_2$ or CO$_2$ emission
at Spitzer wavelengths detectable in our data. In contrast, LkHa 330,  SR 21 and HD135344B show CO in emission in the fundamental 
rovibrational band at 4.7\,$\mu$m \citep{Pontoppidan08, Salyk09}. Given the detection rates of 
molecular emission around regular disks, the current sample of transitional disks is too small to 
draw any firm statistical conclusions. HD135344B is an F star, so may be of sufficiently early type to lack 
molecular emission for the same reasons that the A and B stars lack it.  CoKu Tau/4 is not
a true disk in transition since dust and gas appears to have been cleared out by a stellar companion and not as 
a result of disk evolution \citep{Ireland08}. The chance that the remaining 3 transitional disks do not show H$_2$O emission by
coincidence, given the detection rate of the remaining disks, is only a few percent, but we do not consider this enough
to conclude that no transitional disks will show molecular emission (especially if deeper spectra are acquired). 

The absence of strong molecular emission from these systems is similar to the results of high signal-to-noise IRS spectroscopy of 
the transitional object TW Hya \citep{Najita10}.  The spectrum shows a striking lack of strong emission features of H$_2$O, C$_2$H$_2$, 
and HCN in the 10-20\,$\mu$m region, although weak emission is detected from other molecules (H$_2$, OH, CO$_2$, HCO$^+$, and tentatively CH$_3$).  
Najita et al. describe how the lack of strong molecular emission is consistent with the possibility that the inner disk has been 
cleared by an orbiting giant planet, although chemical and/or excitation effects may be responsible instead. 

A lack of strong molecular emission from transitional disks is consistent with significant
photodestruction due to insufficient shielding from dust in the optically thin inner region of the disks. In this
scenario, CO is still seen in rovibrational transitions due to a combination of survival through self-shielding and the fact that CO has a mechanism for 
UV fluorescent excitation that is efficient also for cold, low density gas. The latter could be the case for SR 21, for which the CO gas 
originates from a ring at 6-7\,AU, distances at which UV fluorescence of CO is likely operating. The CO gas from HD 135344B, on the other hand,
originates mostly from small radii \citep[$< 1\,$AU][]{Pontoppidan08}. 

\subsection{CO$_2$ sources}

\begin{figure}
\centering
\includegraphics[width=8cm]{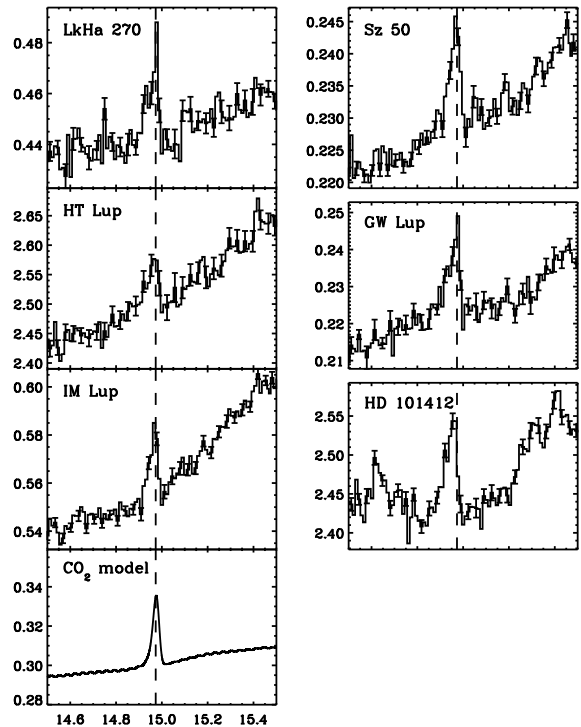}
\caption[]{Detected CO$_2$ emission lines at 14.98\,$\mu$m for sources with no other unambiguous
 molecular emission. For clarity, only every third error bar is plotted. A generic model of 
CO$_2$ emission from a disk, generated using RADLite \citep{Pontoppidan09} is shown
in the lower, left panel.}
\label{CO2only}
\end{figure}

A significant sub-class of disks has been identified, consisting of sources showing strong 
14.98\,$\mu$m CO$_2$ emission from the Q-branch of the fundamental bending mode, 
$v=(0,1,0)\rightarrow (0,0,0)$, but which show no other detectable molecular emission in the 
Spitzer wavelength range. Because the identification with CO$_2$ is based on only one feature, 
it was confirmed that the emission is present in both the A and B nod positions and that 
it is not due to an artifact from the edge of order 13, which begins at 15.08\,$\mu$m. 
Some oxygen-rich asymptotic giant branch (AGB) stars also show strong CO$_2$ 
emission \citep{Justtanont98,Cami00}, but with much higher optical depth leading to strong 
features from combination bands at 13.87 and 16.18\,$\mu$m. These (intrinsically weaker)
bands are absent from protoplanetary disk spectra, suggesting relatively low optical depth of 
the bending mode. The data from the CO$_2$ disks are shown in Fig. \ref{CO2only}, and compared 
to a generic disk model from \citep{Meijerink09} generated using the 
radiative transfer code RADLite \citep{Pontoppidan09}, assuming level populations in local 
thermodynamic equilibrium and a CO$_2$ abundance of $6\times 10^{-8}$ relative to H$_2$. This
abundance is 1-2 orders of magnitude lower than that inferred from chemical models 
\citep{Willacy09}, but given that the critical density for the CO$_2$ bending mode upper state
is high \citep[$10^{10}-10^{12}\,\rm cm^{-3}$, ][]{Castle06}
non-LTE calculations will probably result in higher abundances. Spectra at higher resolution 
should reveal strong R and P branches in the 14-16\,$\mu$m range.  Further chemical 
and radiative transfer modeling is required, not so much to explain the presence of the CO$_2$ 
emission, but why that from water, HCN and C$_2$H$_2$ is absent. Are the abundances of these 
molecules really lower relative to CO$_2$, or is the absence of emission due to small differences 
in the density and temperature structure of the molecular surface layer of these disks causing 
other molecules to be subthermally excited or shielded by dust?

\section{Conclusions}
We have found that a large fraction of protoplanetary disks around low-mass to Solar-type stars have mid- to 
far-IR spectra that are blanketed in emission lines from a wide range of molecules at 
temperatures of 500-1000\,K (Fig. \ref{H2O_rotladder}, \cite{Carr08}, \cite{Salyk08}).
The lines have been shown to be excited in the 0.1-10 AU region of the disk, corresponding 
to the {\it planet-forming zone} \citep{Carr08,Salyk08,Pontoppidan09,Meijerink09}. 
The key conclusion is that such emission appears to be {\it common} in the disks around Sun-like stars. 
It is not a rare or exotic phenomenon, but is rather an unequivocal 
statement that the chemical environment in the planet-forming zone is extremely rich. 
While just a handful of abundant molecules have been identified so far due to the 
relatively low spectral resolution of space-based instruments, the Spitzer results 
demonstrate the concurrent presence of O-, C- and N-chemistry. 
With these ingredients and at the densities and temperatures of the inner disks, 
the chemistry must be highly complex. Given these data, it is the 
expectation that IR studies at higher spectral resolution will reveal many
more molecular species. 

Why have these complex infrared emission spectra not previously been 
seen \citep[e.g.,][]{Meeus01,Kessler06}? Part of the reason is that 
the brighter disks around early type stars that dominated high resolution 
spectral surveys in the past (with the ISO-SWS, for example) do not show 
the same strong emission as seen from T Tauri disks. In fact, disks around 
early type (A and B) stars generally lack molecular emission strong enough 
to be detected by the Spitzer IRS SH modules. It cannot be ruled out that 
some Herbig Ae/Be stars will have molecular emission, as is indicated by 
Figure \ref{coolwater}. One apparently exceptional B star in this sample 
shows emissions due to CO$_2$. Further, OH has been detected at 3\,$\mu$m in 
several Herbig stars \citep{Mandell08}, while rovibrational CO emission is common
in such systems \citep[e.g.,][]{Blake04,Brittain07,vanderplas09}. The 
incidence rate of detectable molecular emission from Herbig stars in
Spitzer spectra appears to be at least 10 times less common than for disks 
around later type stars, however. The reason for this is presently unclear, 
but could be due to a combination of photodestruction of molecules by the 
strong UV fields of the A and B stars, a weakening of lines due to physical 
conditions of the disk, such as a reduced gas-to-small-dust ratio relative 
to T Tauri disks, and masking by a strong infrared continuum. 

Further, this lack of molecular emission from Herbig Ae/Be stars shows the importance of 
extending surveys with Herschel to spectral types of at least G and F. However, the low 
excitation water lines in the Herschel-PACS range are so optically thick that water 
abundances even as low as $10^{-10}$ are expected to produce strong lines 
\citep{Meijerink08, Pontoppidan09}. The mid-infrared lines, in contrast, require
water abundances in the $10^{-4}-10^{-8}$ range, and are therefore important tracers 
of inner disk chemistry. 

The Spitzer data indicate that planetesimals within the snow line generally form in a 
gaseous environment with a high water abundance, assuming that vertical transport
is efficient enough to ensure that the disk surface is representative of the interior within
the midplane snow line. This suggests that the formation of oceans on terrestrial 
planets may not require seeding by a late veneer from the outer reaches of a planetary system, 
such as the late heavy bombardment event, if water can efficiently 
adsorb to grains \citep[e.g.][]{Drake05, Muralidharan08}.
Conversely, the presence of abundant water vapor within the snow line strongly hints at 
a large, unseen reservoir of ice beyond $\sim 1\,$AU. 

The mid-IR bands of water and other molecules are key tracers of planet 
formation. Given their ubiquity, future observations by the JWST, SOFIA and ground-based 
facilities such as the European Extremely Large Telescope (E-ELT), the Thirty Meter 
Telescope (TMT) and the Giant Magellan Telescope (GMT) will have a rich list of targets and 
provide an abundance of constraints for chemical models of inner protoplanetary disks.

\acknowledgments{This work is based on observations made with the Spitzer Space Telescope,
which is operated by the Jet Propulsion Laboratory, California Institute of
Technology under a contract with NASA. Support for this work was provided by NASA.
Support for KMP was provided by NASA through Hubble Fellowship grant \#01201.01 
awarded by the Space Telescope Science Institute, which is operated by the Association of 
Universities for Research in Astronomy, Inc., for NASA, under contract NAS 5-26555. 
Research support for JSC was also provided by 6.1 base funding at the Naval Research Laboratory. 
The authors would like to acknowledge valuable discussions with Uma Gorti, Ilaria Pascucci, 
and Ewine van Dishoeck.}

\bibliographystyle{apj}
\bibliography{ms}

\end{document}